\documentclass[%
10pt,
superscriptaddress,
amsmath,
amssymb,
aps,
twocolumn,
pra,
]{revtex4-2}
\usepackage[colorlinks=true,urlcolor=blue,citecolor=blue,linkcolor=blue]{hyperref}
\usepackage{graphicx}% Include figure files
\usepackage{dcolumn}% Align table columns on decimal point
\usepackage{bm}% bold math
\usepackage{braket}
\usepackage{amsmath}
\usepackage{physics}
\usepackage{lineno}
\usepackage{mathtools,leftindex,tensor,mhchem}
\usepackage{color,soul}

\begin{document}

\preprint{APS/123-QED}

\title{Parity-dependent photon-subtracted two-mode squeezed vacuum states
%Odd and even photon-subtracted two-mode squeezed vacuum states
}

\author{Ananga Mohan Datta}
\affiliation{Humboldt-Universit\"at zu Berlin, Institut f\"ur Physik, AG Theoretische Optik \& Photonik, Newtonstr. 15, 12489 Berlin, Germany}
\affiliation{Okinawa Institute of Science and Technology Graduate University, Onna-son, Okinawa 904-0495, Japan}
\author{Andrea Blanco-Redondo}
\affiliation{CREOL, The College of Optics and Photonics, University of Central Florida, Orlando, Florida 32816, USA}
\author{Kurt Busch}
\affiliation{Humboldt-Universit\"at zu Berlin, Institut f\"ur Physik, AG Theoretische Optik \& Photonik, Newtonstr. 15, 12489 Berlin, Germany}
\affiliation{Max-Born-Institut, Max-Born-Str. 2A, 12489 Berlin, Germany}
\author{Armando Perez-Leija}
\affiliation{Department of Electrical and Computer Engineering, Saint Louis University, St. Louis, Missouri 63103, USA}

\date{\today}% It is always \today, today,

\begin{abstract}
The removal of photons from certain quantum light sources produces so-called photon-subtracted states with enhanced mean photon numbers and intricate quantum correlations. Here, we propose an integrated photon-subtraction scheme that, contrary to previous approaches, is not heralded by photon correlation (coincidence)  measurements. In this way, our technique exploits the ``Welcher Weg'' information problem as it does not provide information about the modes from which the subtracted photons emanated. We show that this lack of information allows the generation of multiphoton states endowed with rich quantum correlations that strongly depend on the parity, evenness or oddness, of the  number of subtracted photons.  
\end{abstract}

\maketitle

%\tableofcontents
%\textit{Introduction.}---
\section{Introduction}
Sculpting the quantum and statistical properties of light is an essential element for the development of quantum-optical information technology \cite{Datta:PRA:2011}.
As a matter of fact, most of such technologies rely on state preparation that requires the engineering of quantum correlations between multiple photons distributed among different modes of the systems in question \cite{DELLANNO200653}.
Admittedly, the generation of quantum correlated multiphoton wavepackets can be achieved using 
linear optical schemes \cite{KnillLaflammeMilburn} in combination with nonlinear light sources \cite{Abouraddy2001}.
In this context, the notion of photon subtraction has emerged as a pragmatic method for the generation of highly correlated multiphoton states \cite{MARCO201041,Carranza:JOSAB:2012,Fan:PRA:2018,Magaña-Loaiza:NQI:2019}. 
Its essence consists in deliberately removing a number of photons from a quantum light source, and counterintuitively this procedure can transform Gaussian states into non-Gaussian ones endowed with enhanced mean photon numbers and appealing nonclassical properties \cite{Zavatta:NJP:2008}.\\
In this context, the non-classicality of zero-photon, single-photon, and multi-photon 
subtracted states from single-mode radiation fields has been the focus of several investigations \cite{Nunn:PRA:2022,Zavatta:NJP:2008,Endo:OE:2023}, while the correlation properties of photon-subtracted two-mode entangled states have also been investigated \cite{Ortiz_QR:2021}. Concurrently, photon-subtraction techniques have been applied to multimode fields, e.g., multimode thermal states \cite{Katamadze:PRA:2020}, two-mode squeezed thermal states \cite{Meng:AP:2020}, and two-mode squeezed vacuum states (TMSVSs) \cite{Thapliyal:PRR:2024, Perina:24}.
Interestingly, the application of photon-subtraction to TMSVSs has recently yielded to the observation of highly-correlated multiphoton states comprising up to ten photons \cite{Magaña-Loaiza:NQI:2019}. This latter achievement, combined with the fact that TMSVSs are the most accessible highly-entangled multiphoton states \cite{Abouraddy2001}, makes photon-subtracted TMSVSs extremely appealing for quantum-enhanced applications, e.g., quantum interferometry \cite{Carranza:JOSAB:2012}, teleportation protocols~\cite{Opatrný:PRA:2000}, quantum illumination \cite{Fan:PRA:2018}, and enhancement of quantum entanglement \cite{Ourjoumtsev:PRL:2007,Bartley:PRA:2013}.

In practice, the photon subtraction scheme applied to TMSVSs, $\ket{r}=\sqrt{1-\vert r\vert^{2}}\sum_{n=0}^{\infty}r^{n}\ket{n}_{a}\ket{n}_{b}$, $r$ being the squeeze parameter, is implemented using two beamsplitters placed in the paths traced out by the modes. Further, the state characterization is done using four photon-number resolving (PNR) detectors, two for counting the reflected photons, that is, the subtracted photons, and two to register the transmitted photons, i.e., the photons left in the initial beams \cite{Magaña-Loaiza:NQI:2019}. In this setting there are two possible photon-subtraction scenarios: i) the symmetric case when the same number of photons is subtracted from each beam, and ii) the asymmetric one when a different number of photons is subtracted. 
Mathematically, both subtraction processes are described by the transformation $\hat{a}^l \hat{b}^m \ket{r} = \mathcal{N}^{-1} \sqrt{1 - |r|^2} \sum_n r^l n! ((n - l)! (n - m)!)^{-1/2} \ket{n - l}_a \ket{n - m}_b$, where the sum runs from $n=l(n=m)$ if $l>m(m<l)$. Here, $\mathcal{N}$ is the normalizing constant, and $\hat{a}$ and $\hat{b}$ are the bosonic annihilation operators for modes $a$ and $b$, respectively.
However, since the use of two PNR detectors to herald photon-subtracted states provides, in principle, precise information about the number of photons subtracted from each beam, this subtraction scheme yields infinite superpositions of photon number states. These are of the form $\alpha\ket{m}_{a}\ket{m}_{b}$ in the symmetric case and $\beta\ket{m}_{a}\ket{n}_{b}$ in the asymmetric case, but not superpositions that combine both types.

In this work, we investigate an alternative photon-subtraction scheme that arises when both modes comprised in a TMSVS source become equally coupled to a common channel in such a way that when measuring the number of photons hopping into this channel, we will know the number of photons being extracted from the modes, but we will not be able to distinguish from which mode the photons arriving at the detector emanated. This lack of information allows the modes comprised in the photon-subtracted states to be inhabited by infinite superpositions of joint photon number states of the type $\alpha_{1}\ket{m}_{a}\ket{n}_{b}+\alpha_{2}\ket{n}_{a}\ket{m}_{b}+\alpha_{3}\ket{m}_{a}\ket{m}_{b}$ or $\beta_{1}\ket{m}_{a}\ket{n}_{b}+\beta_{2}\ket{n}_{a}\ket{m}_{b}$, depending on the parity, evenness or oddness, of the number  of extracted photons. That is, our scheme will allow the coexistence of superpositions of symmetric and asymmetric multiphoton two-mode states with different degree of correlation. \\

To place our findings on solid ground, we propose an experimental feasible scheme based on an integrated waveguide trimer, which can be implemented using current technology, e.g, silicon nanowires \cite{Andrea2018}, or femtosecond-laser-writting techniques \cite{Szameit_2010}. In this latter technique the required couplings can be accomplished by judiciously varying the waveguide separation distance, where, in the weak coupling regime, the coupling coefficients vary exponentially with the separation distance, while the propagation constants can be tuned by varying the waveguide writing velocity, see reference \cite{Szameit_2010} for details.
The waveguide trimer configuration has previously been studied as a versatile platform for the generation and manipulation of multipartite quantum states, including multimode W states \cite{Perez-Leija:PRA:2013, Graefe2014} and three-mode squeezed states \cite{Rojas:PRA:2019}.

Having the waveguide trimer configuration in mind, in what follows we explore the multiphoton quantum correlations exhibited by the emerging photon-subtracted states, and we analyze how the parity of the subtracted photons influence such correlations.

\section{Quantum interference in a waveguide trimer}
\begin{figure}[t!]
    \centering
    \includegraphics[width=0.47\textwidth, height=0.2\textwidth]{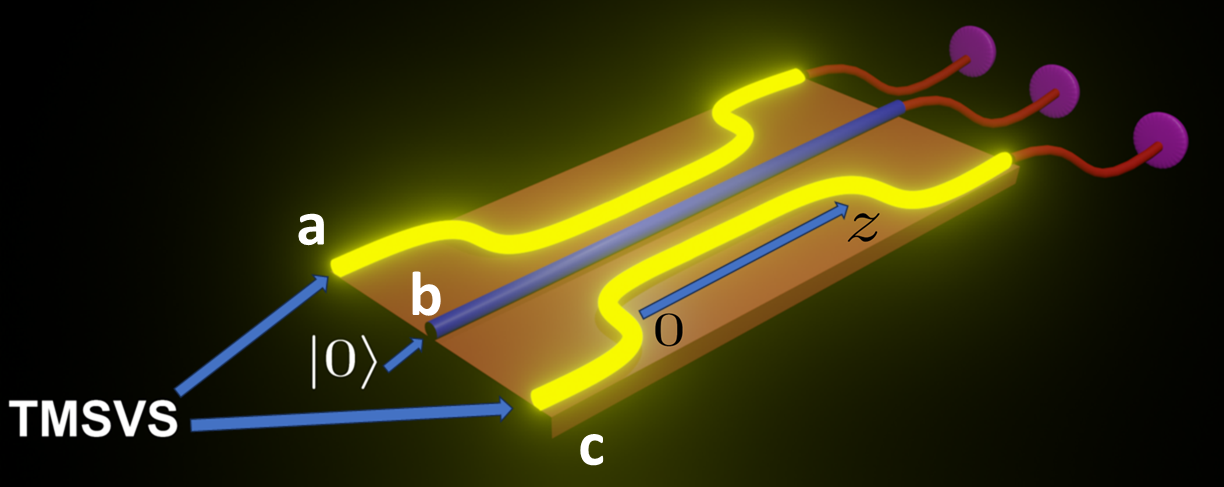}
    \caption{The schematic setup of an integrated waveguide trimer proposed for generating photon-subtracted TMSVS. The modes comprised in the TMSVS source are coupled into the outermost waveguides, while photons are subtracted from the central port $b$ using  a PNR detector. %The resulting photon-subtracted TMSVS are heralded using two PNR detectors at output ports $a$ and $c$.
    }
    \label{fig: Scheme}
    \hrulefill
\end{figure}

Consider a trimer formed by three coupled single-mode waveguides with identical propagation constant $\beta$. The dynamics of the quantized electromagnetic field in such a system is governed by the Heisenberg equations of motion 
\begin{align}\label{eq:couple-mode}
    i\dv{z}\begin{pmatrix}
\hat{a}^{\dagger}(z)\\
\hat{b}^{\dagger}(z)\\
\hat{c}^{\dagger}(z)
\end{pmatrix}=-\begin{pmatrix}
\beta & \kappa & 0\\
\kappa & \beta & \kappa\\
0 & \kappa & \beta
\end{pmatrix}\begin{pmatrix}
\hat{a}^{\dagger}(z)\\
\hat{b}^{\dagger}(z)\\
\hat{c}^{\dagger}(z)
\end{pmatrix},
\end{align}
where $\kappa$ is the coupling constant between adjacent waveguides, $z$ is the propagation coordinate, and $\hat{a}^{\dagger}$, $\hat{b}^{\dagger}$, and $\hat{c}^{\dagger}$ are the bosonic creation operators for waveguides $a$, $b$, and $c$, respectively.
Integration of Eq.~\eqref{eq:couple-mode} yields the input-output relation $\ket{\Psi(z)}=U(z)\ket{\Psi(0)}$, with $\ket{\Psi(z)}=\left(\hat{a}^{\dagger}(z),\hat{b}^{\dagger}(z),\hat{c}^{\dagger}(z)\right)^{T}$, and the evolution matrix
\begin{align}\label{eq:evolution}
U(z)=
    \begin{pmatrix}
\frac{1}{2}+\frac{1}{2}\cos{\Theta} & -\frac{i\sin{\Theta}}{\sqrt{2}} & -\frac{1}{2}+\frac{1}{2}\cos{\Theta}\\
-\frac{i\sin{\Theta}}{\sqrt{2}} & \cos{\Theta} & -\frac{i\sin{\Theta}}{\sqrt{2}}\\
-\frac{1}{2}+\frac{1}{2}\cos{\Theta} & -\frac{i\sin{\Theta}}{\sqrt{2}} & \frac{1}{2}+\frac{1}{2}\cos{\Theta}
\end{pmatrix},
\end{align} where $\Theta=\sqrt{2}\kappa z$. 
We now assume that the modes from a TMSVS source are launched into the two outermost waveguides, while the central waveguide remains in the vacuum state as illustrated in Fig.~\ref{fig: Scheme}. That is, the initial state is $\ket{\psi_{\rm in}}=(1-\vert r\vert ^{2})^{\frac{1}{2}} \sum_{l=0}^{\infty}r^{l}\ket{l}_{a}\ket{0}_{b}\ket{l}_{c}$, and  the photon-subtraction is achieved by monitoring the number of photons hopping into the central waveguide. 
Further, to maximize light transmission at the output of the outermost waveguides while maintaining a feasible probability of photon hopping to the central waveguide, we choose the optimized length of the coupling section such that the ratio between the intensity in the central waveguide $(I_{\rm cen})$ and the intensity in waveguides $a$ and $c$ ($I_{\rm out}$) is $I_{\rm cen}/I_{\rm out}=10/90$, that is,  $z=z_f \approx 0.23/\kappa$. For the details of the calculations, we refer to Appendix \ref{sec:appendix_a}. 

Using the input-output relation, we obtain the output state $\ket{\psi_{\rm out}}=U(z_{f})\ket{\psi_{\rm in}}$, given by
\begin{align}\label{eq:output_state}
\begin{split}
    \ket{\psi_{\rm out}}= & (1-\vert r \vert^{2})^{\frac{1}{2}}\sum_{l=0}^{\infty}\sum_{p_{1},p_{2},p_{3}=0}\sum_{v_{1},v_{2},v_{3}=0}K\\&\ket{v_{1}}_{a}\ket{v_{2}}_{b}\ket{v_{3}}_{c},
\end{split}
\end{align}
where
\begin{align}\label{eq:expression_K}
\begin{split}
    K= & \frac{r^{l}}{p_{1}!p_{2}!p_{3}!}\frac{l!}{(v_{1}-p_{1})!(v_{2}-p_{2})!(v_{3}-p_{3})!}\\& U_{11}^{p_{1}+v_{3}-p_{3}}U_{12}^{v_{2}}U_{13}^{p_{3}+v_{1}-p_{1}}\sqrt{v_{1}!v_{2}!v_{3}!}\delta_{p_{1}+p_{2}+p_{3},l} \\&
    \delta_{v_{1}+v_{2}+v_{3},2l},
\end{split}
\end{align}
which in turn yields the photon-subtracted state 
${\hat{\rho}_{{\rm sub}}}=Tr_{b}(\hat{P}_{b}\ket{\psi_{\rm out}}\bra{\psi_{\rm out}})$,
where ${Tr}_b(.)$ is the partial trace with respect to mode $b$, and $\hat{P}_{b}=:\frac{\eta\hat{n}_{b}}{N!}e^{-\eta\hat{n}_{b}}:$ is the positive operator-valued measure (POVM) operator representing a PNR detector with a quantum efficiency $\eta$. 
Notice, $\hat{P}_{b}$ represents the detection process of $N$ photons in waveguide $b$, the symbol $: :$ implies the normal ordering prescription \cite{Mandel:book:1995}, and $\hat{n}_{b}$ is the photon number operator acting on mode $b$. 
In what follows we assume perfect PNR detectors $(\eta=1)$, at the end we lift this restriction to explore the influence  of minor imperfections in the detection process $(\eta<1)$. For the details of the derivation of the analytical expression for $\hat{\rho}_{\rm{sub}}$ we refer to Appendix \ref{sec:appendix_b}.\\

\begin{figure}[t!]
\centering\includegraphics[width=0.48\textwidth,height=0.4\textwidth] 
    {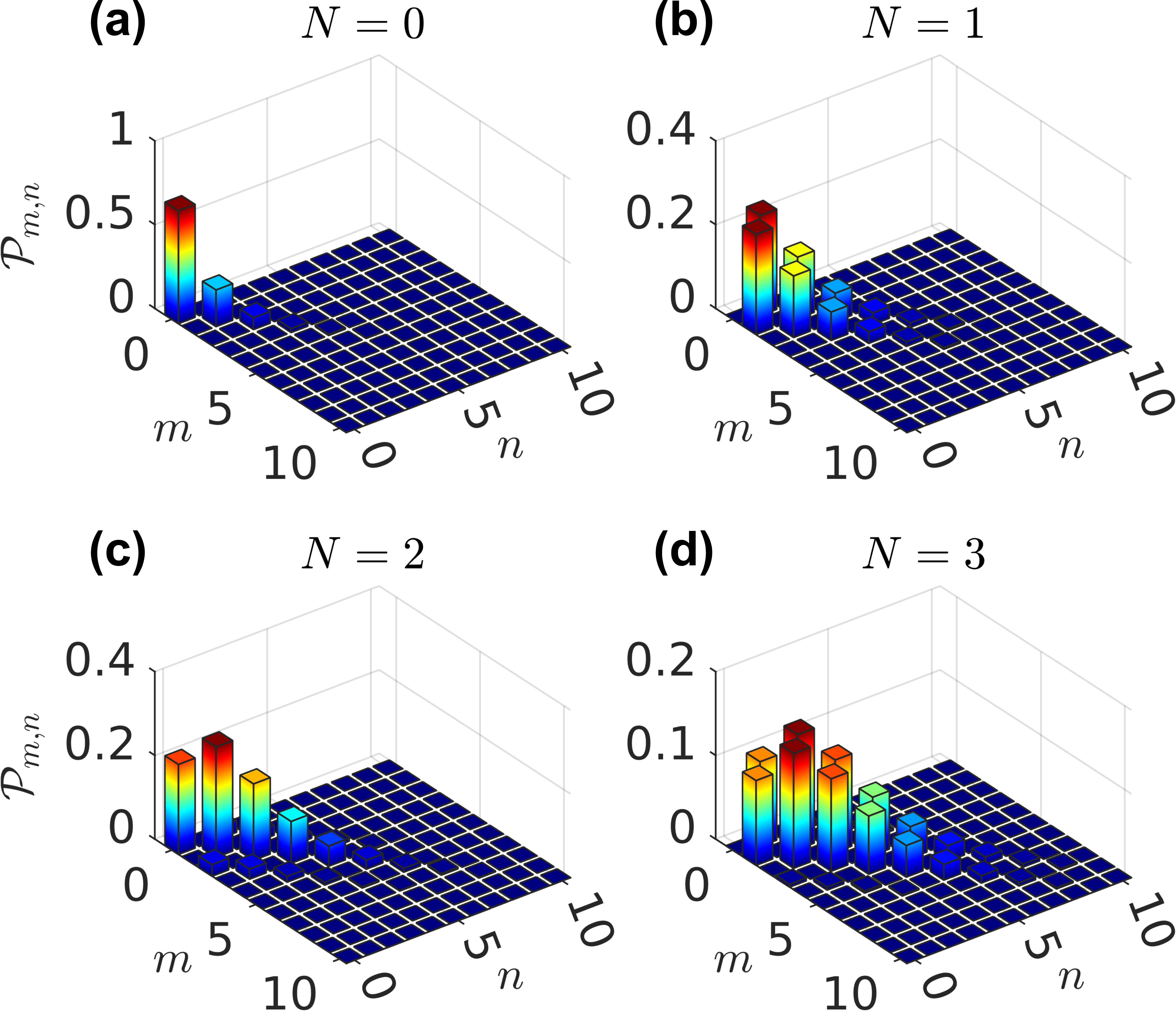}
    \caption{Joint photon-number probability distributions for photon-subtracted TMSVS with $|r|=0.6$, $z=z_{f}$, and detector efficiency $\eta=1$. Figures (a), (b), (c), and (d) illustrate the distributions for subtracting zero ($N=0$), one ($N=1$), two ($N=2$), and three ($N= 3$) photons from the central port, respectively.
}
    \label{fig:Distribution}
    \hrulefill
\end{figure}

To analyze the quantum correlation of the photon-subtracted states, we compute the joint photon-number probability distribution
\begin{align}\label{eq:CorrMat}
\mathcal{P}_{m,n}=\expval{:\frac{(\eta\hat{n}_{a})^{m}}{m!}e^{-\eta\hat{n}_{a}}\otimes \frac{(\eta\hat{n}_{c})^{n}}{n!}e^{-\eta\hat{n}_{c}}:},
\end{align}
which characterizes the probability of detecting $m$ photons in waveguide $a$ and $n$ photons in waveguide $c$, when $N$ photons are detected (subtracted) at waveguide $b$.
Fig.~\ref{fig:Distribution} depicts the computed joint photon-number distribution, wherein zero to three photons have been detected in the central waveguide. 
For the trivial case of zero photon subtraction $(N=0)$, the photon-number distribution contains only diagonal elements 
indicating that most of the time no photons will emerge from the outermost waveguides, Fig.~\ref{fig:Distribution}(a). More interesting scenarios occur when even and odd numbers of photons are subtracted. 

For instances of even photon subtraction ($N$ being even), the resultant state is a superposition of multiphoton states of the form $\alpha_1\left(\ket{m}_{a}\ket{n}_{b}+\ket{n}_{a}\ket{m}_{b}\right)$, and $\alpha_2\ket{m}_{a}\ket{m}_{b}$. The former states result from asymmetric photon subtraction, where a different number of photons is subtracted from the outermost waveguides, while the latter states are due to symmetric photon subtraction in which the same number of photons is being subtracted. 
In other words, detecting an even number of photons in the central waveguide heralds a scenario in which photons are simultaneously subtracted symmetrically and asymmetrically from the outermost modes. However, a close inspection of the joint photon-number distribution reveals that even photon-subtracted states exhibit strong (weak) photon (anti-)correlations as witnessed by the prominent diagonal (small off-diagonal) matrix elements, e.g. Fig.~\ref{fig:Distribution}(c).

\begin{figure}[t!]
\centering\includegraphics[width=0.47\textwidth,height=0.37\textwidth] 
    {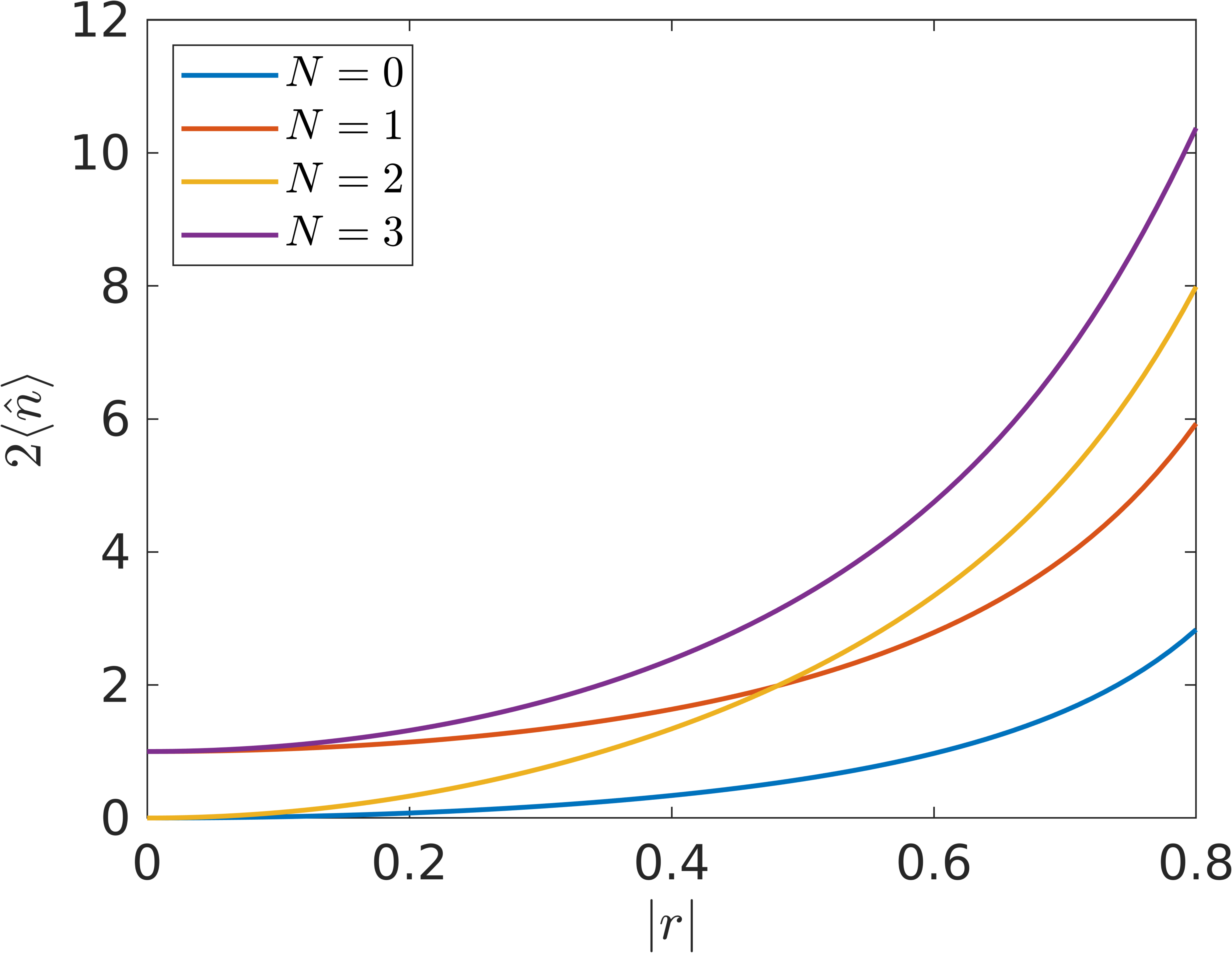}
    \caption{The total mean photon number of a photon-subtracted TMSVS, $2\expval{\hat{n}}$, as a function of $\vert r \vert$ for $N=0,1,2,$ and $3$ at a propagation distance of $z=z_{f}$, assuming a quantum efficiency of $\eta=1$.}
    \label{fig:avg_photons}
    \hrulefill
\end{figure}

Conversely, for odd photon subtraction ($N$ being odd), the resulting states comprise a superposition of states of the form $\alpha_1\left(\ket{m}_{a}\ket{n}_{b}+\ket{n}_{a}\ket{m}_{b}\right)$, rendering a photon number distribution with only off-diagonal elements. This occurs solely due to the asymmetric photon subtraction, again without knowledge about which waveguide the photons are taken from.
The absence of diagonal elements, or equivalently the lack of states $\ket{m}_{a}\ket{m}_{b}$, gives rise to a central nodal line in the distribution with two peaks located on either side of this line, see Figs. \ref{fig:Distribution}(b) and \ref{fig:Distribution}(d).
Significantly, the peaks of the distributions for even and odd photon subtraction show a noticeable shift towards higher photon numbers as the number of subtracted photons $N$ increases, this indicates an overall increase in the average photon number. This increment is qualitatively verified by computing the average photon number in one of the modes, $\expval{\hat{n}}=\sum_{m,n=0}^{\infty}n\mathcal{P}_{m,n}$ which due to the symmetric nature of the joint photon-number distribution gives a total mean photon number $2\expval{\hat{n}}$. In Fig.~\ref{fig:avg_photons}, we present the total average photon number $2\expval{\hat{n}}$ versus the squeeze parameter $\vert r \vert$ for different subtracted number of photons $N$.
Interestingly, for squeezing values $\vert r \vert<0.5$, the average photon number corresponding to odd photon-subtracted states exceeds the one obtained from even photon-subtracted states. Beyond $\vert r \vert=0.5$, the mean photon number for $N=3$ yields a higher mean photon number than all the cases presented here. We would like to point out that this behavior is reminiscent to the one observed for Schrödinger-cat states as discussed in \cite{Dakna:PRA:1997}. Importantly, both symmetric and asymmetric photon subtraction can be advantageous, depending on the target quantum protocol.  For instance, Ref. \cite{Fan:PRA:2018} shows that asymmetric photon subtraction from a TMSVS enhances quantum illumination more effectively than the symmetric approach. Likewise, Ref. \cite{Bartley:PRA:2013} reports that asymmetric subtraction results in a greater entanglement gain compared to symmetric subtraction.
\\
As demonstrated in previous investigations, photon subtracted TMSVSs feature stronger quantum correlations than their TMSVS ``parents'' \cite{Magaña-Loaiza:NQI:2019}. A natural approach to explore such correlations is through the analysis of the matrix of moments \cite{Vogel:PRL:2008}. 
Indeed, theoretical discussions \cite{Sperling:PRA:2013} and experimental demonstrations utilizing PNR detectors \cite{Magaña-Loaiza:NQI:2019} and click detectors \cite{Sperling:PRL:2015} have highlighted nonclassical correlations using the framework of the matrix of moments.
To perform such analysis in the present context, we define the operator $\hat{m}_{a(c)}=\eta\hat{n}_{a(c)}$, such that  
the joint moments are given by $\expval{:\hat{m}_{a}^{u} \otimes  \hat{m}_{c}^{v} :}=\sum_{m,n=0}^{\infty}\mathcal{P}_{m,n}m(m-1)...(m-u+1)n(n-1)...(n-v+1)$, where $u$ and $v$ are positive integers. Importantly, this latter expression allows us to construct a matrix of moments of any order using the joint photon-number distribution $\mathcal{P}_{m,n}$. However, as previously shown \cite{Magaña-Loaiza:NQI:2019}, the second-order matrix of moments suffices to shed light onto the quantum correlations of the generated states
 \begin{align}\label{eq:correlation_matrix}
     M=\begin{pmatrix}
\langle:\hat{m}_{a}^{0}\hat{m}_{c}^{0}:\rangle & \langle:\hat{m}_{a}^{1}\hat{m}_{c}^{0}:\rangle & \langle:\hat{m}_{a}^{0}\hat{m}_{c}^{1}:\rangle\\
\langle:\hat{m}_{a}^{1}\hat{m}_{c}^{0}:\rangle & \langle:\hat{m}_{a}^{2}\hat{m}_{c}^{0}:\rangle & \langle:\hat{m}_{a}^{1}\hat{m}_{c}^{1}:\rangle\\
\langle:\hat{m}_{a}^{0}\hat{m}_{c}^{1}:\rangle & \langle:\hat{m}_{a}^{1}\hat{m}_{c}^{1}:\rangle & \langle:\hat{m}_{a}^{0}\hat{m}_{c}^{2}:\rangle
\end{pmatrix}.
 \end{align}
 \begin{figure}[b!]
\centering\includegraphics[width=0.47\textwidth,height=0.37\textwidth] 
    {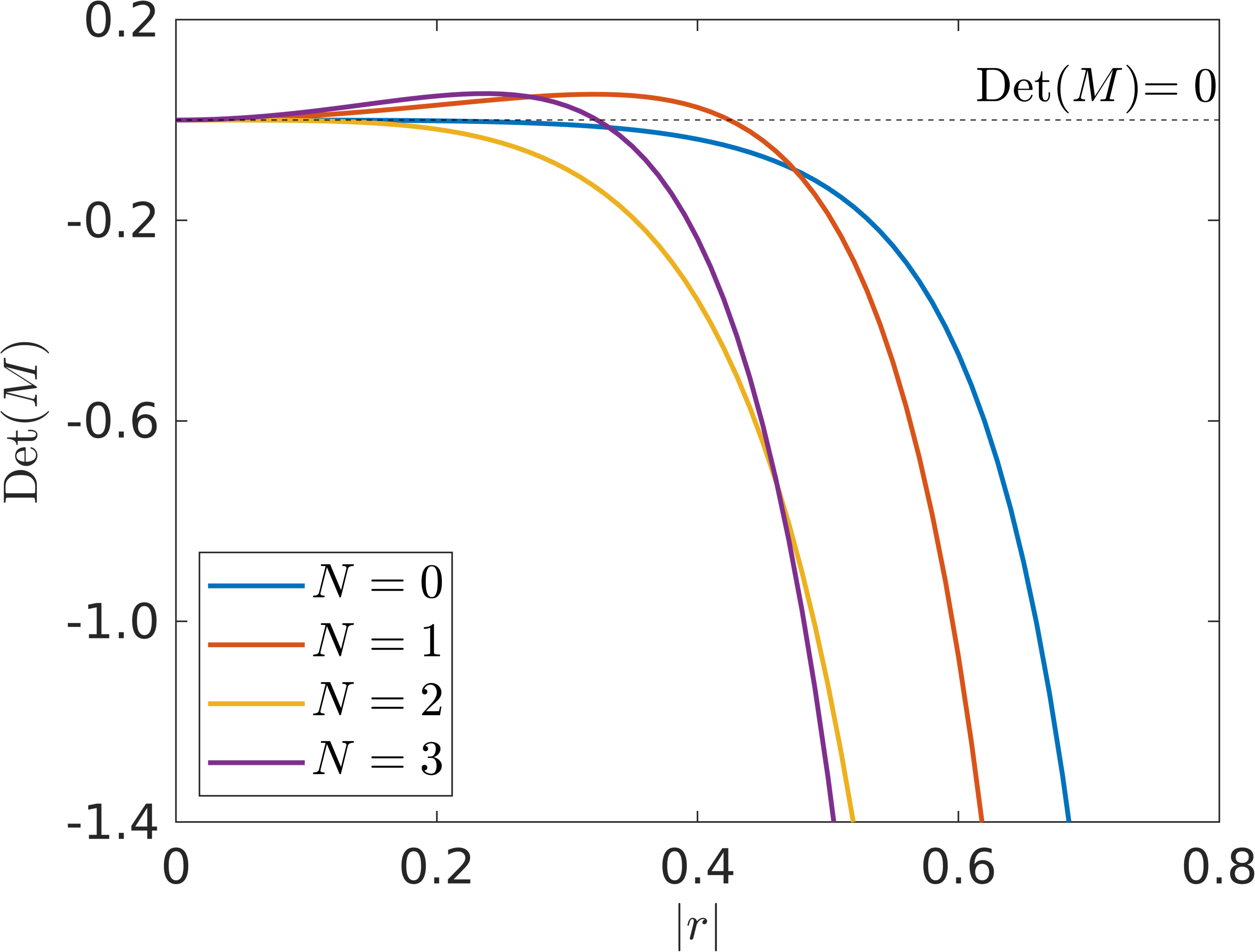}
    \caption{The determinant of the second-order matrix of moments $M$, is examined as a function of $\vert r \vert$ to quantify the correlations between the two output modes of a photon-subtracted TMSVS. This analysis is conducted for values of $N=0,1,2,$ and $3$ at a propagation distance of $z=z_{f}$, with the assumption of a quantum efficiency of $\eta=1$. %The black dotted line represent Det$(M)=0$.
    }
    \label{fig:Correlation_perfect}
    \hrulefill
\end{figure}
Notice, the presence of nonclassical correlations is witnessed whenever Det$(M)<0$. Conversely, Det$(M)>0$ implies classical correlations.
We provide details on computing the second-order moment matrix in Appendix \ref{sec:appendix_d}.
In Fig.~\ref{fig:Correlation_perfect}, Det$(M)$ is plotted against the squeeze parameter $\vert r \vert$ for different values of $N$. 
Notably, the uncertainty regarding which beam the photons are extracted from, leads to intriguing observations about the nature of the correlations.
For $N=1$ our results indicate the existence of classical correlations in the interval $\vert r \vert\le0.5$. Interestingly, this classical interval is reduced as a larger odd number of photons is subtracted, as clearly seen for the case $N=3$ which yields a shorter classical interval $\vert r \vert\le0.4$. 
On the other hand, for $N$ being even, nonclassical correlations emerge in a larger interval $\vert r \vert\in [0.2,1]$. 
These examples suggest that for small squeezing, odd photon subtraction destroys the quantum correlations of the initial TMSVS \cite{Serafini_2005}.
In both scenarios, however, the combination of a TMSVS source with large $\vert r \vert$ and high photon-subtraction, facilitates the production of multiphoton states with enhanced nonclassical correlations. The behavior observed for photon-subtracted TMSVSs with small squeezing parameter is reminiscent to the one reported for two-mode Schrödinger cat states \cite{Gerry:PRA:1995, Hackerv:NP:2019} in which considering even and odd two-mode coherent states with low coherent amplitudes exhibit distinct nonclassical characteristics \cite{Hackerv:NP:2019}.

\begin{figure}[t!]
\centering\includegraphics[width=0.47\textwidth,height=0.37\textwidth] 
    {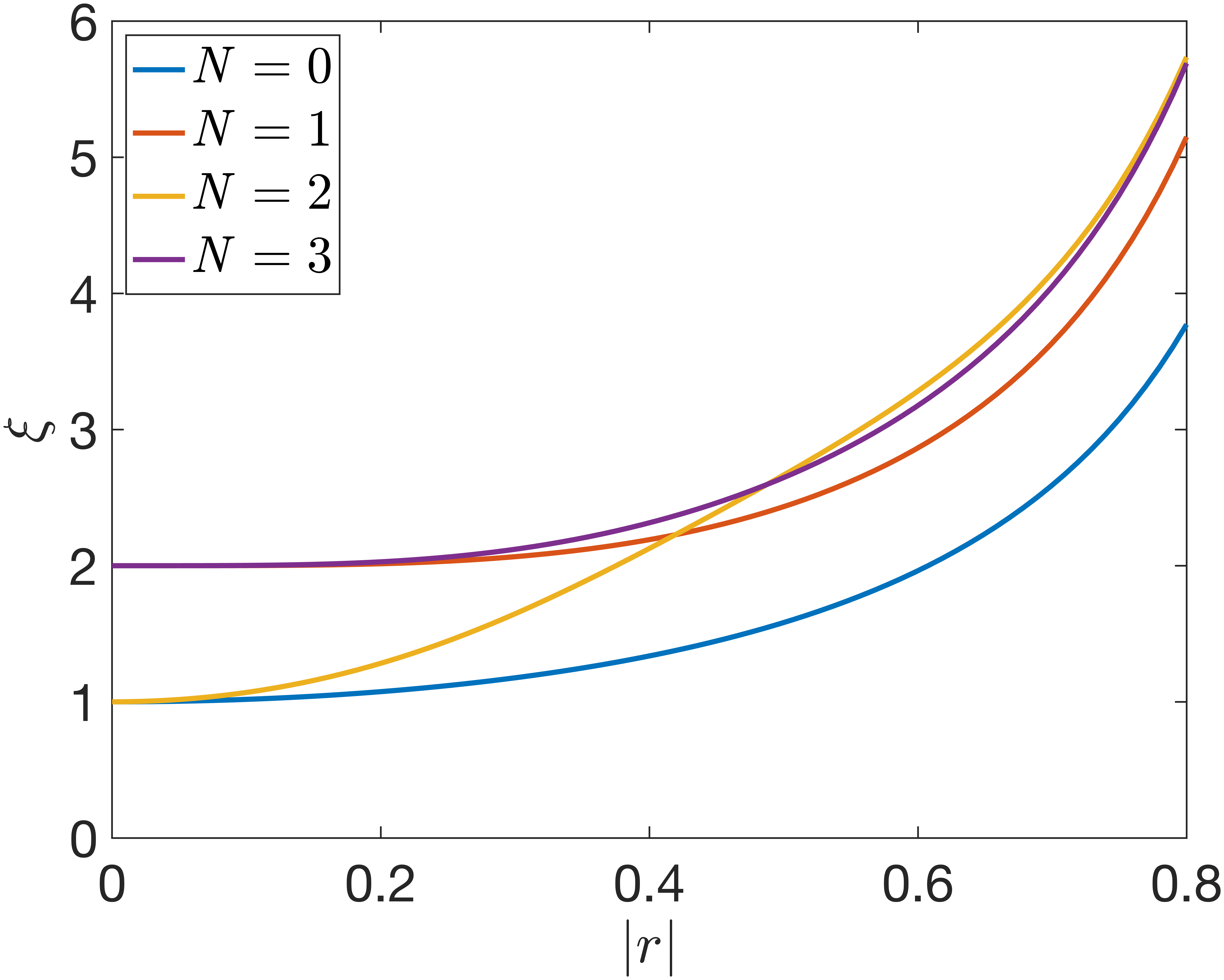}
    \caption{The partcipation ratio $\xi$ is examined as a function of $|r|$ to quantify the degree of entanglement of the photon-subtracted TMSVS. This analysis is performed for photon subtraction numbers $N = 0, 1, 2,$ and $3$, at a fixed propagation distance $z = z_f$, assuming ideal detection with quantum efficiency $\eta = 1$.
    }
    \label{fig:participation_ratio}
    \hrulefill
\end{figure}
To fully characterize the system, we analyze the entanglement of the photon-subtracted two-mode squeezed vacuum state (TMSVS). Photon subtraction is well-known to enhance the entanglement of the parent TMSVSs \cite{Bartley:PRA:2013, Fan:PRA:2018}. To quantify this enhancement, we compute the participation ratio $\xi = 1/\sum_i \lambda_i^2$~\cite{benenti2007tools}, where $\lambda_i$ are the Schmidt coefficients of the bipartite photon-subtracted state. These coefficients are derived from the reduced density matrix of one subsystem (subsystem $( a)$), obtained by partially tracing over the other (subsystem $(c)$)
$\rho_a = \mathrm{Tr}_c\left[ \ket{\psi_{\rm out}}\bra{\psi_{\rm out}} \right] = \sum_i \lambda_i^2 |u_i\rangle \langle u_i|$.
A separable state yields a minimal participation ratio of $(\xi =1)$, while $(\xi >1)$ directly indicates the presence of entanglement~\cite{benenti2007tools}.

Figure \ref{fig:participation_ratio} clearly shows how the participation ratio $\xi$ (our entanglement quantifier) depends on the squeezing parameter $|r|$ for different photon subtraction numbers $(N = 0, 1, 2, 3)$. A key finding is the significant influence of the parity of $N$. Notably, within the low squeezing regime $(|r| \lesssim 0.4)$, odd photon subtraction (specifically, $N=1$ and $N=3$) consistently leads to higher $\xi$ values compared to even cases ($N=0$ and $N=2$). This suggests that subtracting an odd number of photons proves more efficient for enhancing entanglement when the initial state shows weak squeezing.

At this point is worth emphasizing that our theoretical model hinges on the assumption of ideal waveguide propagation, where propagation constants are equal and coupling coefficients are uniform along the propagation coordinate, as describe by Eq.~\eqref{eq:couple-mode}. This is a reasonable premise given the sophistication of current fabrication technologies, such as femtosecond-laser written waveguide platforms, which exhibit low propagation losses of the order $< 0.3 dB/cm$~\cite{Weimann2016, Selim:Science:2025}, and allow for precise control over coupling coefficients and propagation constants. 
These high-quality characteristics lead us to conclude that any imperfections introduced during the state generation process is negligible for our analysis, and any imperfection in the detection scheme can be modeled by assuming detectors with a quantum efficiency parameter $\eta < 1$~\cite{Magaña-Loaiza:NQI:2019}. Hence, before closing, we comment on the effects of using imperfect PNR detectors to herald photon-subtracted states. 

%In our theoretical model, we assume ideal waveguide propagation with equal propagation constants and uniform coupling, as described by Eq.~\eqref{eq:couple-mode}. While this is a simplification, it is justified by the high fabrication quality of femtosecond-laser written waveguide platforms, which typically exhibit low propagation losses (on the order of 0.3~dB/cm) and allow for precise control of propagation constants and coupling coefficients~\cite{Weimann2016, Selim:Science:2025}. These characteristics suggest that imperfections introduced during state generation remain sufficiently small for the purposes of our analysis. 
%Detection inefficiency is modeled separately via the detector quantum efficiency parameter $\eta < 1$~\cite{Magaña-Loaiza:NQI:2019}. 

To evaluate the effect of non-ideal detection, we analyse the determinant of the second-order matrix of moments for $N=0,1,2$, and $3$, considering various values of the parameter $|r|$, and we assume a detector efficiency $\eta=0.8$. From the results presented in Table~\ref{tab:correlations_imperfections}, we note that the distinctive correlation patterns between even and odd photon-subtracted states persist despite imperfections in the detection process.  
\begin{table}[t!]
\centering
\caption{The determinant of the second-order matrix of moments $M$ is calculated for various values of $\vert r \vert$ to examine the impact of imperfections stemming from the PNR detection process. This calculations are conducted for $z=z_{f}$ and $\eta=0.8$.}
\begin{ruledtabular}
\begin{tabular}{ccccc}
$\vert r \vert$ & $N=0$ & $N=1$ & $N=2$ & $N=3$\\
\hline
$0.2$ & -$0.62$ & $1.71$ & -$7.49$ & $8.48$\\
$0.3$ & -$3.79$ & $9.35$ & -$40.67$ & $8.18$\\
\end{tabular}
\label{tab:correlations_imperfections}
\end{ruledtabular}
\end{table}

\section{CONCLUSION}

To conclude, we have proposed and analyzed a waveguide trimer to engineer photon-subtracted TMSVSs, effectively reducing the number of required PNR detectors
to three, compared to the four PNR detectors typically
needed in the conventional photon subtraction process
from twin beams \cite{Magaña-Loaiza:NQI:2019, Thapliyal:PRR:2024, Perina:24}.
Moreover, our scheme requires monitoring only a single output port for photon subtraction, eliminating the need for coincidence detection typically employed in standard twin-beam configurations.
Notably, we observe distinct characteristics in the photon-subtracted state when
retrieving an odd or even number of photons from the
central waveguide. Specifically, we find that the nature
of correlations exhibit a clear dependence on the parity
of the extracted photons, particularly at low values of the
squeeze parameter, resembling the behavior of two-mode
cat states. We have considered the effect of small imperfections in PNR detection on the correlation properties of the photon-subtracted TMSVSs. However, we refrain from a detailed analysis of other experimental imperfections, as such effects are highly dependent on the specific implementation \cite{Magaña-Loaiza:NQI:2019, Sperling:PRL:2015}, waveguide parameters \cite{Szameit_2010, Politi:Science}, and PNR detector characteristics \cite{Provazník:OE:2013}.

Looking forward, the proposed photon-subtraction scheme can be applied to other interesting two-mode quantum states, e.g., NOON states.
Our protocol could potentially open up new avenues for investigating multiphoton process in integrated photonics platforms \cite{Anaya-Contreras_2021}. 
Introducing detuning might also provide a valuable way to study more nuanced quantum correlations in such systems \cite{Datta:24}. In a broader context, this work represents a step toward scalable and stable quantum-optical implementations within integrated photonics, where such approaches may support the development of future quantum technologies \cite{Wang:NP:2020, Laurent:PRXQ:2024}.
The feasibility of our approach is further reinforced by recent technological advances in integrated quantum photonics, including quantum state purification protocols \cite{Wang2025} and simplified continuous-variable quantum key distribution implementations \cite{Zhao:22}.
In this context, our work holds practical significance in various correlation-based schemes. For instance, the creation of multiphoton states with controlled correlations through photon subtraction could notably enhance applications in quantum illumination \cite{Fan:PRA:2018}, and improve robustness of Boson sampling in integrated photonics circuits \cite{Spring:Science:2013}.
Finally, as the unveiled quantum correlations exhibited by the present multiphoton photon-subtracted states has not been discussed in previous investigations \cite{Thapliyal:PRR:2024, Carranza:JOSAB:2012, Opatrný:PRA:2000, Fan:PRA:2018, Ourjoumtsev:PRL:2007, Magaña-Loaiza:NQI:2019}, our results may open the door to new explorations on the generation of more complex multiphoton light states.

%Additionally, previous studies have showcased the implementation of Boson sampling through the photon subtraction method \cite{Olson:PRA:2015}. Hence, our device could be useful to improve the robustness of Boson sampling in integrated photonics circuits \cite{Spring:Science:2013}.

\textbf{Funding.} A.M.D., K.B, and A.P.L. acknowledge financial support by the Leibniz Association within
the Leibniz Collaborative Excellence Program (Project ID K266/2019, On-chip Laser-written
Photonic Circuits for Classical and Quantum Applications (LAPTON)). 
A.B.-R. acknowledges support by the National Science Foundation (NSF) (Award No. 2328993).
A.M.D. and K.B. acknowledge funding by the German Research Foundation (DFG) in the
framework of the Collaborative Research Center (CRC) 1375 (Project ID 398816777 - Project A06)

\textbf{Disclosures.} The authors declare no conflicts of interest.

\textbf{Data availability.} Data underlying the results presented in this paper are not publicly available at this time but may be obtained from the authors upon reasonable request.

\appendix\markboth{Appendix}{Appendix}
\renewcommand{\thesection}{\Alph{section}}
\numberwithin{equation}{section}
\section{Calculating the propagation distance $z_{f}$}
\label{sec:appendix_a}
In this Appendix, we determine the propagation distance $z=z_{f}$ required to establish a
specific ratio of light intensity between the output ports of the two outermost
waveguides, denoted by $a$ and $c$, compared to that of the central waveguide $b$ (see Fig.~\ref{fig: Scheme}).
To begin, we envision a scenario where both input ports $a$ and $c$ of the 
waveguide trimer, as depicted in Fig.~\ref{fig: Scheme}, receive simultaneous
excitation by photons. In this context, the input state can be represented as
\begin{align}
     |\phi(0)\rangle=\frac{1}{\sqrt{2}}\begin{pmatrix}
1\\
0\\
1
\end{pmatrix}.
 \end{align}
 To trace the evolution of the system, we find the output state at a 
 propagation distance $z$ within the trimer, which is computed using the evolution operator in Eq.~\eqref{eq:evolution}
\begin{align}
    |\phi(z)\rangle =\hat{U}(z)\frac{1}{\sqrt{2}}\begin{pmatrix}
    1\\
    0\\
    1
    \end{pmatrix}
    =\begin{pmatrix}
        \frac{\cos({\sqrt{2}z\kappa})}{\sqrt{2}}\\
        -i\sin{\sqrt{2}z\kappa}\\
        \frac{\cos({\sqrt{2}z\kappa})}{\sqrt{2}}
    \end{pmatrix}.
\end{align}
This leads us to derive the output intensity in the two outermost waveguides $I_{\rm out}$ and the central waveguide $I_{\rm cen}$, expressed as
\begin{align}
    I_{\rm out}=\cos^{2}{\sqrt{2}z\kappa},
\end{align}
\begin{align}
    I_{\rm cen}=\sin^{2}{\sqrt{2}z\kappa}.
\end{align}
With this in hand, we can deduce the ratio of intensity at the output ports
of the outermost waveguides to that of the central waveguide, which is given by
\begin{align}\label{Intensity_ratio}
    \frac{I_{\rm cen}}{I_{\rm out}}=\tan^{2}(\sqrt{2}z\kappa).
\end{align}

By establishing the target ratio $I_{\rm cen}/I_{\rm out}$ as $10/90$, we can
determine the associated propagation distance
$z=z_{f}$ through Eq.~\eqref{Intensity_ratio}, yielding an approximate
value of $z_{f} \approx \frac{0.23}{\kappa}$.
\section{Analytical expression of photon-subtracted TMSVS}
\label{sec:appendix_b}
In this Appendix, we provide a derivation of the analytical expression for the photon-subtracted TMSVS.
 We initiate our analysis via the input state
 \begin{align}
 \begin{split}
     \ket{\psi_{\rm in}}&=(1-\vert r\vert^{2})^{\frac{1}{2}} \sum_{n=0}^{\infty}r^{l}\ket{l}_{a}\ket{0}_{b}\ket{l}_{c}\\
     &=(1-\vert r \vert ^{2})^{\frac{1}{2}} \sum_{n=0}^{\infty}r^{l}\frac{(\hat{a}^{\dagger})^{l}}{\sqrt{l!}}
     \frac{(\hat{c}^{\dagger})^{l}}{\sqrt{l!}}\ket{0}_{a}\ket{0}_{b}\ket{0}_{c},
\end{split}
 \end{align}
where $\hat{a}^{\dagger}$ and $\hat{c}^{\dagger}$ are creation operator acting
on mode $a$ and $c$, respectively.
Next, we proceed by employing the bosonic operator transformation 
to compute the output state $\ket{\psi_{\rm out}}$ 
 \begin{align}\label{eq:5}
 \begin{split}
     \ket{\psi_{\rm out}}=&(1-\vert r \vert^{2})^{\frac{1}{2}}\sum_{l=0}^{\infty}\frac{r^{l}}{l!}(U_{11}\hat{a}^{\dagger}+U_{12}\hat{b}^{\dagger}+U_{13}\hat{c}^{\dagger})^{l}\\&(U_{31}\hat{a}^{\dagger}+U_{32}\hat{b}^{\dagger}+U_{33}\hat{c}^{\dagger})^{l}\ket{0}_{a}\ket{0}_{b}\ket{0}_{c}.
\end{split}
 \end{align}
Utilizing the multinomial theorem, we obtain
\begin{align}\label{eq:output_state_appendix}
\begin{split}
    \ket{\psi_{\rm out}}= & (1-\vert r \vert^{2})^{\frac{1}{2}}\sum_{l=0}^{\infty}\sum_{p_{1},p_{2},p_{3}=0}\sum_{v_{1},v_{2},v_{3}=0}K\\&\ket{v_{1}}_{a}\ket{v_{2}}_{b}\ket{v_{3}}_{c},
\end{split}
\end{align}
where
\begin{align}\label{eq:expression_K}
\begin{split}
    K= & \frac{r^{l}}{p_{1}!p_{2}!p_{3}!}\frac{l!}{(v_{1}-p_{1})!(v_{2}-p_{2})!(v_{3}-p_{3})!}\\& U_{11}^{p_{1}+v_{3}-p_{3}}U_{12}^{v_{2}}U_{13}^{p_{3}+v_{1}-p_{1}}\sqrt{v_{1}!v_{2}!v_{3}!}\delta_{p_{1}+p_{2}+p_{3},l} \\&
    \delta_{v_{1}+v_{2}+v_{3},2l}.
\end{split}
\end{align}
It is important to note that we utilize the Hermitian nature of the evolution 
matrix in Eq.~\eqref{eq:evolution}, where $U_{11}=U_{33}$ and $U_{12}=U_{32}$.
Moving forward, the photon-subtracted state for detecting $N$ photons in the central waveguide by a photon-number resolving (PNR) detector with efficiency $\eta$ is computed as 
\begin{align}\label{eq:photon_subtracted_state}
    \hat{\rho}_{\rm sub}=Tr_{b}(\hat{P}_{b}|\psi_{\rm out}\rangle\langle\psi_{\rm out}|)
\end{align}
where $\hat{P}_{b}$ is known as the POVM operator for a PNR detector,
given by
\begin{align}\label{eq: POVM_operator}
\begin{split}
    \hat{P}_{b}&=:\frac{(\eta\hat{n}_{b})^{N}}{N!}e^{-\eta\hat{n}_{b}}:\\
    &=\sum_{j=N}^{\infty}\begin{pmatrix}j\\N \end{pmatrix}\eta^{N}(1-\eta)^{j-N}|j\rangle_{b}\langle j|_{b}.
\end{split}
\end{align}
Using Eqs.~\eqref{eq:output_state_appendix}, ~\eqref{eq:photon_subtracted_state}, and ~\eqref{eq: POVM_operator}, we derive the analytical expression for the photon-subtracted TMSVS
\begin{align}
\begin{split}
        \hat{\rho}_{\rm sub}= & (1-|r|^{2})\sum_{l=0}^{\infty}\sum_{l^{\prime}=0}^{\infty}\sum_{p_{1},p_{2},p_{3}=0}\sum_{p_{1}^{\prime},p_{2}^{\prime},p_{3}^{\prime}=0}\sum_{v_{1},v_{3}=0}
        \sum_{v^{\prime}_{1},v^{\prime}_{3}=0}  \\&
        \sum_{j=N}\begin{pmatrix}
            j\\N
        \end{pmatrix}\eta^{N}(1-\eta)^{j-N}S \ket{v_{1}}_{a}\ket{v_{3}}_{c} \\ & 
        \leftindex_{c}{\bra{v_{3}^{\prime}}}\leftindex_{a}{\bra{v_{1}^{\prime}}}\delta_{p_{1}+p_{2}+p_{3},l} \delta_{p_{1}^{\prime}+p_{2}^{\prime}+p_{3}^{\prime},l}\delta_{v_{1}+j+v_{3}, 2l}
        \\ & \delta_{v_{1}^{\prime}+j+v_{3}^{\prime}, 2l^{\prime}}
\end{split}
\end{align}
where
\begin{align}
\begin{split} 
    S=& \frac{r^{l}}{p_{1}!p_{2}!p_{3}!}\frac{l!}{(v_{1}-p_{1})!(j-p_{2})!(v_{3}-p_{3})!}     
    \frac{r^{l^{\prime}}}{p^{\prime}_{1}!p^{\prime}_{2}!p^{\prime}_{3}!} \\ &
    \frac{l^{\prime}!}{(v_{1}^{\prime}-p^{\prime}_{1})!(j-p^{\prime}_{2})!(v_{3}^{\prime}-p_{3}^{\prime})!} U_{11}^{p_{1}+v_{3}-p_{3}}   (U^{*}_{11})^{p_{1}^{\prime}+v_{3}^{\prime}-p_{3}^{\prime}} 
    \\& |U_{12}|^{2j}U_{13}^{p_{3}+v_{1}-p_{1}}(U^{*}_{13})^{p_{3}^{\prime}+v^{\prime}_{1}-p_{1}^{\prime}}j!\sqrt{v_{1}!v_{3}!v_{1}^{\prime}!v_{3}^{\prime}!}.
\end{split}
\end{align}

\section{Photon-number distribution}
\label{sec:appendix_c}
Here, we derive the photon-number distribution of the photon-subtracted 
TMSVS measured in ports $a$ and $c$. The probability to measure $m$ and $n$ photons 
in ports $a$ and $c$, respectively, using PNR detectors of quantum efficiency $\eta$, is given by
\begin{align}
    \begin{split}
        \mathcal{P}_{m,n}&=\expval{:\frac{(\eta\hat{n}_{a})^{m}}{m!}e^{-\eta\hat{n}_{a}}\otimes \frac{(\eta\hat{n}_{c})^{n}}{n!}e^{-\eta\hat{n}_{c}}:}\\&=\frac{P_{N}(m,n)}{\sum_{m,n=0}P_{N}(m,n)},\\
    \end{split}
\end{align}
where
\begin{align}
    \begin{split}
        P_{N}(m,n)= & (1-|r|^{2})\sum_{l=0}^{\infty}\sum_{p_{1},p_{2},p_{3}=0}\sum_{p_{1}^{\prime},p_{2}^{\prime},p_{3}^{\prime}=0}\sum_{j=N}\begin{pmatrix}
            j\\N
        \end{pmatrix}\eta^{N}\\&(1-\eta)^{j-N}\sum_{i=m}\begin{pmatrix}
            i\\m
        \end{pmatrix}\eta^{m}(1-\eta)^{i-m}\sum_{k=n}\begin{pmatrix}
            k\\n
        \end{pmatrix}\\ & \eta^{n}(1-\eta)^{k-n}  Q\delta_{p_{1}+p_{2}+p_{3},l}\delta_{p_{1}^{\prime}+p_{2}^{\prime}+p_{3}^{\prime},l}\\& \delta_{v_{1}+v_{2}+v_{3}, 2l}l,
    \end{split}
\end{align}
and 
\begin{align}
\begin{split} 
    Q= & \frac{r^{2l}}{p_{1}!p_{2}!p_{3}!}\frac{(l!)^{2}}{(v_{1}-p_{1})!(j-p_{2})!(v_{3}-p_{3})!}     
    \frac{1}{p^{\prime}_{1}!p^{\prime}_{2}!p^{\prime}_{3}!} \\ &
    \frac{1}{(v_{1}-p^{\prime}_{1})!(j-p^{\prime}_{2})!(v_{3}-p_{3}^{\prime})!} U_{11}^{p_{1}+v_{3}-p_{3}}   (U^{*}_{11})^{p_{1}^{\prime}+v_{3}-p_{3}^{\prime}} 
    \\& |U_{12}|^{2j}U_{13}^{p_{3}+v_{1}-p_{1}}(U^{*}_{13})^{p_{3}^{\prime}+v_{1}-p_{1}^{\prime}}j!v_{1}!v_{3}.
\end{split}
\end{align}

\section{Second-order matrix of moments}
\label{sec:appendix_d}
In this Appendix, we calculate the second-order matrix of moments to measure the correlation. 
Each element in the matrix in  Eq.~\eqref{eq:correlation_matrix} can be expressed in terms of joint photon-number distribution $\mathcal{P}_{m,n}$ as follows:

\begin{align}
    \langle:\hat{m}_{a}^{1}\hat{m}_{c}^{0}:\rangle=
    \langle:\hat{m}_{a}^{1}:\rangle=
    \sum_{m=0}^{\infty}\sum_{n=0}^{\infty}m \mathcal{P}_{m,n},
\end{align}

\begin{align}
    \langle:\hat{m}_{a}^{0}\hat{m}_{c}^{1}:\rangle=
    \langle:\hat{m}_{c}^{1}:\rangle=
    \sum_{m=0}^{\infty}\sum_{n=0}^{\infty}n \mathcal{P}_{m,n},
\end{align}

\begin{align}
    \langle:\hat{m}_{a}^{1}\hat{m}_{a}^{1}:\rangle=
    \langle:\hat{m}_{a}^{2}:\rangle=
    \sum_{m=0}^{\infty}\sum_{n=0}^{\infty}m(m-1) \mathcal{P}_{m,n},
\end{align}

\begin{align}
    \langle:\hat{m}_{c}^{1}\hat{m}_{c}^{1}:\rangle=
    \langle:\hat{m}_{c}^{2}:\rangle=
    \sum_{m=0}^{\infty}\sum_{n=0}^{\infty}n(n-1) \mathcal{P}_{m,n},
\end{align}

\begin{align}
    \langle:\hat{m}_{a}^{1}\hat{m}_{c}^{1}:\rangle=
    \sum_{m=0}^{\infty}\sum_{n=0}^{\infty}m n \mathcal{P}_{m,n}.
\end{align}
To obtain the matrix of moments expressed in relation 
to the joint photon-number distribution, we direct 
readers to the supplementary section of reference ~\cite{Magaña-Loaiza:NQI:2019}.

\bibliography{apssamp}% Produces the bibliography via BibTeX.

\begin{thebibliography}{44}%
\makeatletter
\providecommand \@ifxundefined [1]{%
 \@ifx{#1\undefined}
}%
\providecommand \@ifnum [1]{%
 \ifnum #1\expandafter \@firstoftwo
 \else \expandafter \@secondoftwo
 \fi
}%
\providecommand \@ifx [1]{%
 \ifx #1\expandafter \@firstoftwo
 \else \expandafter \@secondoftwo
 \fi
}%
\providecommand \natexlab [1]{#1}%
\providecommand \enquote  [1]{``#1''}%
\providecommand \bibnamefont  [1]{#1}%
\providecommand \bibfnamefont [1]{#1}%
\providecommand \citenamefont [1]{#1}%
\providecommand \href@noop [0]{\@secondoftwo}%
\providecommand \href [0]{\begingroup \@sanitize@url \@href}%
\providecommand \@href[1]{\@@startlink{#1}\@@href}%
\providecommand \@@href[1]{\endgroup#1\@@endlink}%
\providecommand \@sanitize@url [0]{\catcode `\\12\catcode `\$12\catcode `\&12\catcode `\#12\catcode `\^12\catcode `\_12\catcode `\%12\relax}%
\providecommand \@@startlink[1]{}%
\providecommand \@@endlink[0]{}%
\providecommand \url  [0]{\begingroup\@sanitize@url \@url }%
\providecommand \@url [1]{\endgroup\@href {#1}{\urlprefix }}%
\providecommand \urlprefix  [0]{URL }%
\providecommand \Eprint [0]{\href }%
\providecommand \doibase [0]{https://doi.org/}%
\providecommand \selectlanguage [0]{\@gobble}%
\providecommand \bibinfo  [0]{\@secondoftwo}%
\providecommand \bibfield  [0]{\@secondoftwo}%
\providecommand \translation [1]{[#1]}%
\providecommand \BibitemOpen [0]{}%
\providecommand \bibitemStop [0]{}%
\providecommand \bibitemNoStop [0]{.\EOS\space}%
\providecommand \EOS [0]{\spacefactor3000\relax}%
\providecommand \BibitemShut  [1]{\csname bibitem#1\endcsname}%
\let\auto@bib@innerbib\@empty
%</preamble>
\bibitem [{\citenamefont {Datta}\ \emph {et~al.}(2011)\citenamefont {Datta}, \citenamefont {Zhang}, \citenamefont {Thomas-Peter}, \citenamefont {Dorner}, \citenamefont {Smith},\ and\ \citenamefont {Walmsley}}]{Datta:PRA:2011}%
  \BibitemOpen
  \bibfield  {author} {\bibinfo {author} {\bibfnamefont {A.}~\bibnamefont {Datta}}, \bibinfo {author} {\bibfnamefont {L.}~\bibnamefont {Zhang}}, \bibinfo {author} {\bibfnamefont {N.}~\bibnamefont {Thomas-Peter}}, \bibinfo {author} {\bibfnamefont {U.}~\bibnamefont {Dorner}}, \bibinfo {author} {\bibfnamefont {B.~J.}\ \bibnamefont {Smith}},\ and\ \bibinfo {author} {\bibfnamefont {I.~A.}\ \bibnamefont {Walmsley}},\ }\bibfield  {title} {\bibinfo {title} {Quantum metrology with imperfect states and detectors},\ }\href {https://doi.org/10.1103/PhysRevA.83.063836} {\bibfield  {journal} {\bibinfo  {journal} {Phys. Rev. A}\ }\textbf {\bibinfo {volume} {83}},\ \bibinfo {pages} {063836} (\bibinfo {year} {2011})}\BibitemShut {NoStop}%
\bibitem [{\citenamefont {Anno}\ \emph {et~al.}(2006)\citenamefont {Anno}, \citenamefont {Siena},\ and\ \citenamefont {Illuminati}}]{DELLANNO200653}%
  \BibitemOpen
  \bibfield  {author} {\bibinfo {author} {\bibfnamefont {F.~D.}\ \bibnamefont {Anno}}, \bibinfo {author} {\bibfnamefont {S.~D.}\ \bibnamefont {Siena}},\ and\ \bibinfo {author} {\bibfnamefont {F.}~\bibnamefont {Illuminati}},\ }\bibfield  {title} {\bibinfo {title} {Multiphoton quantum optics and quantum state engineering},\ }\href {https://doi.org/https://doi.org/10.1016/j.physrep.2006.01.004} {\bibfield  {journal} {\bibinfo  {journal} {Physics Reports}\ }\textbf {\bibinfo {volume} {428}},\ \bibinfo {pages} {53} (\bibinfo {year} {2006})}\BibitemShut {NoStop}%
\bibitem [{\citenamefont {Knill}\ \emph {et~al.}(2001)\citenamefont {Knill}, \citenamefont {Laflamme},\ and\ \citenamefont {Milburn}}]{KnillLaflammeMilburn}%
  \BibitemOpen
  \bibfield  {author} {\bibinfo {author} {\bibfnamefont {E.}~\bibnamefont {Knill}}, \bibinfo {author} {\bibfnamefont {R.}~\bibnamefont {Laflamme}},\ and\ \bibinfo {author} {\bibfnamefont {G.~J.}\ \bibnamefont {Milburn}},\ }\bibfield  {title} {\bibinfo {title} {A scheme for efficient quantum computation with linear optics},\ }\href {https://doi.org/10.1038/35051009} {\bibfield  {journal} {\bibinfo  {journal} {Nature}\ }\textbf {\bibinfo {volume} {409}},\ \bibinfo {pages} {46} (\bibinfo {year} {2001})}\BibitemShut {NoStop}%
\bibitem [{\citenamefont {Abouraddy}\ \emph {et~al.}(2001)\citenamefont {Abouraddy}, \citenamefont {Saleh}, \citenamefont {Sergienko},\ and\ \citenamefont {Teich}}]{Abouraddy2001}%
  \BibitemOpen
  \bibfield  {author} {\bibinfo {author} {\bibfnamefont {A.~F.}\ \bibnamefont {Abouraddy}}, \bibinfo {author} {\bibfnamefont {B.~E.~A.}\ \bibnamefont {Saleh}}, \bibinfo {author} {\bibfnamefont {A.~V.}\ \bibnamefont {Sergienko}},\ and\ \bibinfo {author} {\bibfnamefont {M.~C.}\ \bibnamefont {Teich}},\ }\bibfield  {title} {\bibinfo {title} {Role of entanglement in two-photon imaging},\ }\href {https://doi.org/10.1103/PhysRevLett.87.123602} {\bibfield  {journal} {\bibinfo  {journal} {Phys. Rev. Lett.}\ }\textbf {\bibinfo {volume} {87}},\ \bibinfo {pages} {123602} (\bibinfo {year} {2001})}\BibitemShut {NoStop}%
\bibitem [{\citenamefont {Marco}\ and\ \citenamefont {Alessandro}(2010)}]{MARCO201041}%
  \BibitemOpen
  \bibfield  {author} {\bibinfo {author} {\bibfnamefont {B.}~\bibnamefont {Marco}}\ and\ \bibinfo {author} {\bibfnamefont {Z.}~\bibnamefont {Alessandro}},\ }\bibfield  {title} {\bibinfo {title} {Chapter 2 - manipulating light states by single-photon addition and subtraction}\ }(\bibinfo  {publisher} {Elsevier},\ \bibinfo {year} {2010})\ pp.\ \bibinfo {pages} {41--83}\BibitemShut {NoStop}%
\bibitem [{\citenamefont {Carranza}\ and\ \citenamefont {Gerry}(2012)}]{Carranza:JOSAB:2012}%
  \BibitemOpen
  \bibfield  {author} {\bibinfo {author} {\bibfnamefont {R.}~\bibnamefont {Carranza}}\ and\ \bibinfo {author} {\bibfnamefont {C.~C.}\ \bibnamefont {Gerry}},\ }\bibfield  {title} {\bibinfo {title} {Photon-subtracted two-mode squeezed vacuum states and applications to quantum optical interferometry},\ }\href {https://doi.org/10.1364/JOSAB.29.002581} {\bibfield  {journal} {\bibinfo  {journal} {J. Opt. Soc. Am. B}\ }\textbf {\bibinfo {volume} {29}},\ \bibinfo {pages} {2581} (\bibinfo {year} {2012})}\BibitemShut {NoStop}%
\bibitem [{\citenamefont {Fan}\ and\ \citenamefont {Zubairy}(2018)}]{Fan:PRA:2018}%
  \BibitemOpen
  \bibfield  {author} {\bibinfo {author} {\bibfnamefont {L.}~\bibnamefont {Fan}}\ and\ \bibinfo {author} {\bibfnamefont {M.~S.}\ \bibnamefont {Zubairy}},\ }\bibfield  {title} {\bibinfo {title} {Quantum illumination using non-gaussian states generated by photon subtraction and photon addition},\ }\href {https://doi.org/10.1103/PhysRevA.98.012319} {\bibfield  {journal} {\bibinfo  {journal} {Phys. Rev. A}\ }\textbf {\bibinfo {volume} {98}},\ \bibinfo {pages} {012319} (\bibinfo {year} {2018})}\BibitemShut {NoStop}%
\bibitem [{\citenamefont {Magaña-Loaiza}\ \emph {et~al.}(2019)\citenamefont {Magaña-Loaiza}, \citenamefont {de~J.~León-Montiel}, \citenamefont {Perez-Leija}, \citenamefont {U’Ren}, \citenamefont {You}, \citenamefont {Busch}, \citenamefont {Lita}, \citenamefont {Nam}, \citenamefont {Mirin},\ and\ \citenamefont {Gerrits}}]{Magaña-Loaiza:NQI:2019}%
  \BibitemOpen
  \bibfield  {author} {\bibinfo {author} {\bibfnamefont {O.~S.}\ \bibnamefont {Magaña-Loaiza}}, \bibinfo {author} {\bibfnamefont {R.}~\bibnamefont {de~J.~León-Montiel}}, \bibinfo {author} {\bibfnamefont {A.}~\bibnamefont {Perez-Leija}}, \bibinfo {author} {\bibfnamefont {A.~B.}\ \bibnamefont {U’Ren}}, \bibinfo {author} {\bibfnamefont {C.}~\bibnamefont {You}}, \bibinfo {author} {\bibfnamefont {K.}~\bibnamefont {Busch}}, \bibinfo {author} {\bibfnamefont {A.~E.}\ \bibnamefont {Lita}}, \bibinfo {author} {\bibfnamefont {S.~W.}\ \bibnamefont {Nam}}, \bibinfo {author} {\bibfnamefont {R.~P.}\ \bibnamefont {Mirin}},\ and\ \bibinfo {author} {\bibfnamefont {T.}~\bibnamefont {Gerrits}},\ }\bibfield  {title} {\bibinfo {title} {Multiphoton quantum-state engineering using conditional measurements},\ }\href {https://doi.org/10.1038/s41534-019-0195-2} {\bibfield  {journal} {\bibinfo  {journal} {npj Quantum Inf}\ }\textbf {\bibinfo {volume} {5}},\ \bibinfo {pages} {80} (\bibinfo {year} {2019})}\BibitemShut {NoStop}%
\bibitem [{\citenamefont {Zavatta}\ \emph {et~al.}(2008)\citenamefont {Zavatta}, \citenamefont {Parigi}, \citenamefont {Kim},\ and\ \citenamefont {Bellini}}]{Zavatta:NJP:2008}%
  \BibitemOpen
  \bibfield  {author} {\bibinfo {author} {\bibfnamefont {A.}~\bibnamefont {Zavatta}}, \bibinfo {author} {\bibfnamefont {V.}~\bibnamefont {Parigi}}, \bibinfo {author} {\bibfnamefont {M.~S.}\ \bibnamefont {Kim}},\ and\ \bibinfo {author} {\bibfnamefont {M.}~\bibnamefont {Bellini}},\ }\bibfield  {title} {\bibinfo {title} {Subtracting photons from arbitrary light fields: experimental test of coherent state invariance by single-photon annihilation},\ }\href {https://doi.org/10.1088/1367-2630/10/12/123006} {\bibfield  {journal} {\bibinfo  {journal} {New J. Phys.}\ }\textbf {\bibinfo {volume} {10}},\ \bibinfo {pages} {123006} (\bibinfo {year} {2008})}\BibitemShut {NoStop}%
\bibitem [{\citenamefont {Nunn}\ \emph {et~al.}(2022)\citenamefont {Nunn}, \citenamefont {Franson},\ and\ \citenamefont {Pittman}}]{Nunn:PRA:2022}%
  \BibitemOpen
  \bibfield  {author} {\bibinfo {author} {\bibfnamefont {C.~M.}\ \bibnamefont {Nunn}}, \bibinfo {author} {\bibfnamefont {J.~D.}\ \bibnamefont {Franson}},\ and\ \bibinfo {author} {\bibfnamefont {T.~B.}\ \bibnamefont {Pittman}},\ }\bibfield  {title} {\bibinfo {title} {Modifying quantum optical states by zero-photon subtraction},\ }\href {https://doi.org/10.1103/PhysRevA.105.033702} {\bibfield  {journal} {\bibinfo  {journal} {Phys. Rev. A}\ }\textbf {\bibinfo {volume} {105}},\ \bibinfo {pages} {033702} (\bibinfo {year} {2022})}\BibitemShut {NoStop}%
\bibitem [{\citenamefont {Endo}\ \emph {et~al.}(2023)\citenamefont {Endo}, \citenamefont {He}, \citenamefont {Sonoyama}, \citenamefont {Takahashi}, \citenamefont {Kashiwazaki}, \citenamefont {Umeki}, \citenamefont {Takasu}, \citenamefont {Hattori}, \citenamefont {Fukuda}, \citenamefont {Fukui}, \citenamefont {Takase}, \citenamefont {Asavanant}, \citenamefont {Marek}, \citenamefont {Filip},\ and\ \citenamefont {Furusawa}}]{Endo:OE:2023}%
  \BibitemOpen
  \bibfield  {author} {\bibinfo {author} {\bibfnamefont {M.}~\bibnamefont {Endo}}, \bibinfo {author} {\bibfnamefont {R.}~\bibnamefont {He}}, \bibinfo {author} {\bibfnamefont {T.}~\bibnamefont {Sonoyama}}, \bibinfo {author} {\bibfnamefont {K.}~\bibnamefont {Takahashi}}, \bibinfo {author} {\bibfnamefont {T.}~\bibnamefont {Kashiwazaki}}, \bibinfo {author} {\bibfnamefont {T.}~\bibnamefont {Umeki}}, \bibinfo {author} {\bibfnamefont {S.}~\bibnamefont {Takasu}}, \bibinfo {author} {\bibfnamefont {K.}~\bibnamefont {Hattori}}, \bibinfo {author} {\bibfnamefont {D.}~\bibnamefont {Fukuda}}, \bibinfo {author} {\bibfnamefont {K.}~\bibnamefont {Fukui}}, \bibinfo {author} {\bibfnamefont {K.}~\bibnamefont {Takase}}, \bibinfo {author} {\bibfnamefont {W.}~\bibnamefont {Asavanant}}, \bibinfo {author} {\bibfnamefont {P.}~\bibnamefont {Marek}}, \bibinfo {author} {\bibfnamefont {R.}~\bibnamefont {Filip}},\ and\ \bibinfo {author} {\bibfnamefont {A.}~\bibnamefont {Furusawa}},\ }\bibfield  {title} {\bibinfo {title} {Non-gaussian
  quantum state generation by multi-photon subtraction at the telecommunication wavelength},\ }\href {https://doi.org/10.1364/OE.486270} {\bibfield  {journal} {\bibinfo  {journal} {Opt. Express}\ }\textbf {\bibinfo {volume} {31}},\ \bibinfo {pages} {12865} (\bibinfo {year} {2023})}\BibitemShut {NoStop}%
\bibitem [{\citenamefont {Rosas-Ortiz}\ and\ \citenamefont {Zelaya}(2021)}]{Ortiz_QR:2021}%
  \BibitemOpen
  \bibfield  {author} {\bibinfo {author} {\bibfnamefont {O.}~\bibnamefont {Rosas-Ortiz}}\ and\ \bibinfo {author} {\bibfnamefont {K.}~\bibnamefont {Zelaya}},\ }\bibfield  {title} {\bibinfo {title} {Theory of photon subtraction for two-mode entangled light beams},\ }\href {https://doi.org/10.3390/quantum3030033} {\bibfield  {journal} {\bibinfo  {journal} {Quantum Reports}\ }\textbf {\bibinfo {volume} {3}},\ \bibinfo {pages} {500} (\bibinfo {year} {2021})}\BibitemShut {NoStop}%
\bibitem [{\citenamefont {Katamadze}\ \emph {et~al.}(2020)\citenamefont {Katamadze}, \citenamefont {Avosopiants}, \citenamefont {Bogdanova}, \citenamefont {Bogdanov},\ and\ \citenamefont {Kulik}}]{Katamadze:PRA:2020}%
  \BibitemOpen
  \bibfield  {author} {\bibinfo {author} {\bibfnamefont {K.~G.}\ \bibnamefont {Katamadze}}, \bibinfo {author} {\bibfnamefont {G.~V.}\ \bibnamefont {Avosopiants}}, \bibinfo {author} {\bibfnamefont {N.~A.}\ \bibnamefont {Bogdanova}}, \bibinfo {author} {\bibfnamefont {Y.~I.}\ \bibnamefont {Bogdanov}},\ and\ \bibinfo {author} {\bibfnamefont {S.~P.}\ \bibnamefont {Kulik}},\ }\bibfield  {title} {\bibinfo {title} {Multimode thermal states with multiphoton subtraction: Study of the photon-number distribution in the selected subsystem},\ }\href {https://doi.org/10.1103/PhysRevA.101.013811} {\bibfield  {journal} {\bibinfo  {journal} {Phys. Rev. A}\ }\textbf {\bibinfo {volume} {\href{}{101}}},\ \bibinfo {pages} {013811} (\bibinfo {year} {2020})}\BibitemShut {NoStop}%
\bibitem [{\citenamefont {Meng}\ \emph {et~al.}(2020)\citenamefont {Meng}, \citenamefont {Li}, \citenamefont {Wang}, \citenamefont {Zhang}, \citenamefont {Zhang}, \citenamefont {Yang},\ and\ \citenamefont {Liang}}]{Meng:AP:2020}%
  \BibitemOpen
  \bibfield  {author} {\bibinfo {author} {\bibfnamefont {X.-G.}\ \bibnamefont {Meng}}, \bibinfo {author} {\bibfnamefont {K.-C.}\ \bibnamefont {Li}}, \bibinfo {author} {\bibfnamefont {J.-S.}\ \bibnamefont {Wang}}, \bibinfo {author} {\bibfnamefont {X.-Y.}\ \bibnamefont {Zhang}}, \bibinfo {author} {\bibfnamefont {Z.-T.}\ \bibnamefont {Zhang}}, \bibinfo {author} {\bibfnamefont {Z.-S.}\ \bibnamefont {Yang}},\ and\ \bibinfo {author} {\bibfnamefont {B.-L.}\ \bibnamefont {Liang}},\ }\bibfield  {title} {\bibinfo {title} {Continuous-variable entanglement and wigner-functionnegativity via adding or subtracting photons},\ }\href {https://doi.org/10.1002/andp.201900585} {\bibfield  {journal} {\bibinfo  {journal} {Ann. Phys.}\ }\textbf {\bibinfo {volume} {532}},\ \bibinfo {pages} {1900585} (\bibinfo {year} {2020})}\BibitemShut {NoStop}%
\bibitem [{\citenamefont {Thapliyal}\ \emph {et~al.}(2024)\citenamefont {Thapliyal}, \citenamefont {Peřina}, \citenamefont {Jr.}, \citenamefont {Michálek},\ and\ \citenamefont {Machulka}}]{Thapliyal:PRR:2024}%
  \BibitemOpen
  \bibfield  {author} {\bibinfo {author} {\bibfnamefont {K.}~\bibnamefont {Thapliyal}}, \bibinfo {author} {\bibfnamefont {J.}~\bibnamefont {Peřina}}, \bibinfo {author} {\bibfnamefont {O.~H.}\ \bibnamefont {Jr.}}, \bibinfo {author} {\bibfnamefont {V.}~\bibnamefont {Michálek}},\ and\ \bibinfo {author} {\bibfnamefont {R.}~\bibnamefont {Machulka}},\ }\bibfield  {title} {\bibinfo {title} {Experimental characterization of multimode photon-subtracted twin beams},\ }\href {https://doi.org/10.1103/PhysRevResearch.6.013065} {\bibfield  {journal} {\bibinfo  {journal} {Phys. Rev. Research}\ }\textbf {\bibinfo {volume} {6}},\ \bibinfo {pages} {013065} (\bibinfo {year} {2024})}\BibitemShut {NoStop}%
\bibitem [{\citenamefont {Pe\v{r}ina}\ \emph {et~al.}(2024)\citenamefont {Pe\v{r}ina}, \citenamefont {Thapliyal}, \citenamefont {Haderka}, \citenamefont {Mich\'{a}lek},\ and\ \citenamefont {Machulka}}]{Perina:24}%
  \BibitemOpen
  \bibfield  {author} {\bibinfo {author} {\bibfnamefont {J.}~\bibnamefont {Pe\v{r}ina}}, \bibinfo {author} {\bibfnamefont {K.}~\bibnamefont {Thapliyal}}, \bibinfo {author} {\bibfnamefont {O.}~\bibnamefont {Haderka}}, \bibinfo {author} {\bibfnamefont {V.}~\bibnamefont {Mich\'{a}lek}},\ and\ \bibinfo {author} {\bibfnamefont {R.}~\bibnamefont {Machulka}},\ }\bibfield  {title} {\bibinfo {title} {Sub-poissonian twin beams},\ }\href {https://doi.org/10.1364/OPTICAQ.509228} {\bibfield  {journal} {\bibinfo  {journal} {Optica Quantum}\ }\textbf {\bibinfo {volume} {2}},\ \bibinfo {pages} {148} (\bibinfo {year} {2024})}\BibitemShut {NoStop}%
\bibitem [{\citenamefont {Opatrný}\ \emph {et~al.}(2000)\citenamefont {Opatrný}, \citenamefont {Kurizki},\ and\ \citenamefont {Welsch}}]{Opatrný:PRA:2000}%
  \BibitemOpen
  \bibfield  {author} {\bibinfo {author} {\bibfnamefont {T.}~\bibnamefont {Opatrný}}, \bibinfo {author} {\bibfnamefont {G.}~\bibnamefont {Kurizki}},\ and\ \bibinfo {author} {\bibfnamefont {D.-G.}\ \bibnamefont {Welsch}},\ }\bibfield  {title} {\bibinfo {title} {Improvement on teleportation of continuous variables by photon subtraction via conditional measurement},\ }\href {https://doi.org/10.1103/PhysRevA.61.032302} {\bibfield  {journal} {\bibinfo  {journal} {Phys. Rev. A}\ }\textbf {\bibinfo {volume} {61}},\ \bibinfo {pages} {032302} (\bibinfo {year} {2000})}\BibitemShut {NoStop}%
\bibitem [{\citenamefont {Ourjoumtsev}\ \emph {et~al.}(2007)\citenamefont {Ourjoumtsev}, \citenamefont {Dantan}, \citenamefont {Tualle-Brouri},\ and\ \citenamefont {Grangier}}]{Ourjoumtsev:PRL:2007}%
  \BibitemOpen
  \bibfield  {author} {\bibinfo {author} {\bibfnamefont {A.}~\bibnamefont {Ourjoumtsev}}, \bibinfo {author} {\bibfnamefont {A.}~\bibnamefont {Dantan}}, \bibinfo {author} {\bibfnamefont {R.}~\bibnamefont {Tualle-Brouri}},\ and\ \bibinfo {author} {\bibfnamefont {P.}~\bibnamefont {Grangier}},\ }\bibfield  {title} {\bibinfo {title} {Increasing entanglement between gaussian states by coherent photon subtraction},\ }\href {https://doi.org/10.1103/PhysRevLett.98.030502} {\bibfield  {journal} {\bibinfo  {journal} {Phys. Rev. Lett.}\ }\textbf {\bibinfo {volume} {98}},\ \bibinfo {pages} {030502} (\bibinfo {year} {2007})}\BibitemShut {NoStop}%
\bibitem [{\citenamefont {Bartley}\ \emph {et~al.}(2013)\citenamefont {Bartley}, \citenamefont {Crowley}, \citenamefont {Datta}, \citenamefont {Nunn}, \citenamefont {Zhang},\ and\ \citenamefont {Walmsley}}]{Bartley:PRA:2013}%
  \BibitemOpen
  \bibfield  {author} {\bibinfo {author} {\bibfnamefont {T.~J.}\ \bibnamefont {Bartley}}, \bibinfo {author} {\bibfnamefont {P.~J.~D.}\ \bibnamefont {Crowley}}, \bibinfo {author} {\bibfnamefont {A.}~\bibnamefont {Datta}}, \bibinfo {author} {\bibfnamefont {J.}~\bibnamefont {Nunn}}, \bibinfo {author} {\bibfnamefont {L.}~\bibnamefont {Zhang}},\ and\ \bibinfo {author} {\bibfnamefont {I.}~\bibnamefont {Walmsley}},\ }\bibfield  {title} {\bibinfo {title} {Strategies for enhancing quantum entanglement by local photon subtraction},\ }\href {https://doi.org/10.1103/PhysRevA.87.022313} {\bibfield  {journal} {\bibinfo  {journal} {Phys. Rev. A}\ }\textbf {\bibinfo {volume} {87}},\ \bibinfo {pages} {022313} (\bibinfo {year} {2013})}\BibitemShut {NoStop}%
\bibitem [{\citenamefont {Blanco-Redondo}\ \emph {et~al.}(2018)\citenamefont {Blanco-Redondo}, \citenamefont {Bell}, \citenamefont {Oren}, \citenamefont {Eggleton},\ and\ \citenamefont {Segev}}]{Andrea2018}%
  \BibitemOpen
  \bibfield  {author} {\bibinfo {author} {\bibfnamefont {A.}~\bibnamefont {Blanco-Redondo}}, \bibinfo {author} {\bibfnamefont {B.}~\bibnamefont {Bell}}, \bibinfo {author} {\bibfnamefont {D.}~\bibnamefont {Oren}}, \bibinfo {author} {\bibfnamefont {B.~J.}\ \bibnamefont {Eggleton}},\ and\ \bibinfo {author} {\bibfnamefont {M.}~\bibnamefont {Segev}},\ }\bibfield  {title} {\bibinfo {title} {Topological protection of biphoton states},\ }\href {https://doi.org/10.1126/science.aau4296} {\bibfield  {journal} {\bibinfo  {journal} {Science}\ }\textbf {\bibinfo {volume} {362}},\ \bibinfo {pages} {568} (\bibinfo {year} {2018})}\BibitemShut {NoStop}%
\bibitem [{\citenamefont {Szameit}\ and\ \citenamefont {Nolte}(2010)}]{Szameit_2010}%
  \BibitemOpen
  \bibfield  {author} {\bibinfo {author} {\bibfnamefont {A.}~\bibnamefont {Szameit}}\ and\ \bibinfo {author} {\bibfnamefont {S.}~\bibnamefont {Nolte}},\ }\bibfield  {title} {\bibinfo {title} {Discrete optics in femtosecond-laser-written photonic structures},\ }\href {https://doi.org/10.1088/0953-4075/43/16/163001} {\bibfield  {journal} {\bibinfo  {journal} {Journal of Physics B: Atomic, Molecular and Optical Physics}\ }\textbf {\bibinfo {volume} {43}},\ \bibinfo {pages} {163001} (\bibinfo {year} {2010})}\BibitemShut {NoStop}%
\bibitem [{\citenamefont {Perez-Leija}\ \emph {et~al.}(2013)\citenamefont {Perez-Leija}, \citenamefont {Hernandez-Herrejon}, \citenamefont {Moya-Cessa}, \citenamefont {Szameit},\ and\ \citenamefont {Christodoulides}}]{Perez-Leija:PRA:2013}%
  \BibitemOpen
  \bibfield  {author} {\bibinfo {author} {\bibfnamefont {A.}~\bibnamefont {Perez-Leija}}, \bibinfo {author} {\bibfnamefont {J.~C.}\ \bibnamefont {Hernandez-Herrejon}}, \bibinfo {author} {\bibfnamefont {H.}~\bibnamefont {Moya-Cessa}}, \bibinfo {author} {\bibfnamefont {A.}~\bibnamefont {Szameit}},\ and\ \bibinfo {author} {\bibfnamefont {D.~N.}\ \bibnamefont {Christodoulides}},\ }\bibfield  {title} {\bibinfo {title} {Generating photon-encoded $w$ states in multiport waveguide-array systems},\ }\href {https://doi.org/10.1103/PhysRevA.87.013842} {\bibfield  {journal} {\bibinfo  {journal} {Phys. Rev. A}\ }\textbf {\bibinfo {volume} {87}},\ \bibinfo {pages} {013842} (\bibinfo {year} {2013})}\BibitemShut {NoStop}%
\bibitem [{\citenamefont {Gr{\"a}fe}\ \emph {et~al.}(2014)\citenamefont {Gr{\"a}fe}, \citenamefont {Heilmann}, \citenamefont {Perez-Leija}, \citenamefont {Keil}, \citenamefont {Dreisow}, \citenamefont {Heinrich}, \citenamefont {Moya-Cessa}, \citenamefont {Nolte}, \citenamefont {Christodoulides},\ and\ \citenamefont {Szameit}}]{Graefe2014}%
  \BibitemOpen
  \bibfield  {author} {\bibinfo {author} {\bibfnamefont {M.}~\bibnamefont {Gr{\"a}fe}}, \bibinfo {author} {\bibfnamefont {R.}~\bibnamefont {Heilmann}}, \bibinfo {author} {\bibfnamefont {A.}~\bibnamefont {Perez-Leija}}, \bibinfo {author} {\bibfnamefont {R.}~\bibnamefont {Keil}}, \bibinfo {author} {\bibfnamefont {F.}~\bibnamefont {Dreisow}}, \bibinfo {author} {\bibfnamefont {M.}~\bibnamefont {Heinrich}}, \bibinfo {author} {\bibfnamefont {H.}~\bibnamefont {Moya-Cessa}}, \bibinfo {author} {\bibfnamefont {S.}~\bibnamefont {Nolte}}, \bibinfo {author} {\bibfnamefont {D.~N.}\ \bibnamefont {Christodoulides}},\ and\ \bibinfo {author} {\bibfnamefont {A.}~\bibnamefont {Szameit}},\ }\bibfield  {title} {\bibinfo {title} {On-chip generation of high-order single-photon w-states},\ }\href {https://doi.org/10.1038/nphoton.2014.204} {\bibfield  {journal} {\bibinfo  {journal} {Nature Photon}\ }\textbf {\bibinfo {volume} {8}},\ \bibinfo {pages} {791} (\bibinfo {year} {2014})}\BibitemShut {NoStop}%
\bibitem [{\citenamefont {Rojas-Rojas}\ \emph {et~al.}(2019)\citenamefont {Rojas-Rojas}, \citenamefont {Barriga}, \citenamefont {Mu\~noz}, \citenamefont {Solano},\ and\ \citenamefont {Hermann-Avigliano}}]{Rojas:PRA:2019}%
  \BibitemOpen
  \bibfield  {author} {\bibinfo {author} {\bibfnamefont {S.}~\bibnamefont {Rojas-Rojas}}, \bibinfo {author} {\bibfnamefont {E.}~\bibnamefont {Barriga}}, \bibinfo {author} {\bibfnamefont {C.}~\bibnamefont {Mu\~noz}}, \bibinfo {author} {\bibfnamefont {P.}~\bibnamefont {Solano}},\ and\ \bibinfo {author} {\bibfnamefont {C.}~\bibnamefont {Hermann-Avigliano}},\ }\bibfield  {title} {\bibinfo {title} {Manipulation of multimode squeezing in a coupled waveguide array},\ }\href {https://doi.org/10.1103/PhysRevA.100.023841} {\bibfield  {journal} {\bibinfo  {journal} {Phys. Rev. A}\ }\textbf {\bibinfo {volume} {100}},\ \bibinfo {pages} {023841} (\bibinfo {year} {2019})}\BibitemShut {NoStop}%
\bibitem [{\citenamefont {Mandel}\ and\ \citenamefont {Wolf}(1995)}]{Mandel:book:1995}%
  \BibitemOpen
  \bibfield  {author} {\bibinfo {author} {\bibfnamefont {L.}~\bibnamefont {Mandel}}\ and\ \bibinfo {author} {\bibfnamefont {E.}~\bibnamefont {Wolf}},\ }\href@noop {} {\emph {\bibinfo {title} {Optical Coherence and Quantum Optics}}}\ (\bibinfo  {publisher} {Cambridge University Press},\ \bibinfo {year} {1995})\BibitemShut {NoStop}%
\bibitem [{\citenamefont {Dakna}\ \emph {et~al.}(1997)\citenamefont {Dakna}, \citenamefont {Anhut}, \citenamefont {Opatrný}, \citenamefont {Knöll},\ and\ \citenamefont {Welsch}}]{Dakna:PRA:1997}%
  \BibitemOpen
  \bibfield  {author} {\bibinfo {author} {\bibfnamefont {M.}~\bibnamefont {Dakna}}, \bibinfo {author} {\bibfnamefont {T.}~\bibnamefont {Anhut}}, \bibinfo {author} {\bibfnamefont {T.}~\bibnamefont {Opatrný}}, \bibinfo {author} {\bibfnamefont {L.}~\bibnamefont {Knöll}},\ and\ \bibinfo {author} {\bibfnamefont {D.-G.}\ \bibnamefont {Welsch}},\ }\bibfield  {title} {\bibinfo {title} {Generating schrödinger-cat-like states by means of conditional measurements on a beam splitter},\ }\href {https://doi.org/10.1103/PhysRevA.55.3184} {\bibfield  {journal} {\bibinfo  {journal} {Phys. Rev. A}\ }\textbf {\bibinfo {volume} {55}},\ \bibinfo {pages} {3184} (\bibinfo {year} {1997})}\BibitemShut {NoStop}%
\bibitem [{\citenamefont {Vogel}(2008)}]{Vogel:PRL:2008}%
  \BibitemOpen
  \bibfield  {author} {\bibinfo {author} {\bibfnamefont {W.}~\bibnamefont {Vogel}},\ }\bibfield  {title} {\bibinfo {title} {Nonclassical correlation properties of radiation fields},\ }\href {https://doi.org/10.1103/PhysRevLett.100.013605} {\bibfield  {journal} {\bibinfo  {journal} {Phys. Rev. Lett.}\ }\textbf {\bibinfo {volume} {100}},\ \bibinfo {pages} {013605} (\bibinfo {year} {2008})}\BibitemShut {NoStop}%
\bibitem [{\citenamefont {Sperling}\ \emph {et~al.}(2013)\citenamefont {Sperling}, \citenamefont {Vogel},\ and\ \citenamefont {Agarwal}}]{Sperling:PRA:2013}%
  \BibitemOpen
  \bibfield  {author} {\bibinfo {author} {\bibfnamefont {J.}~\bibnamefont {Sperling}}, \bibinfo {author} {\bibfnamefont {W.}~\bibnamefont {Vogel}},\ and\ \bibinfo {author} {\bibfnamefont {G.~S.}\ \bibnamefont {Agarwal}},\ }\bibfield  {title} {\bibinfo {title} {Correlation measurements with on-off detectors},\ }\href {https://doi.org/10.1103/PhysRevA.88.043821} {\bibfield  {journal} {\bibinfo  {journal} {Phys. Rev. A}\ }\textbf {\bibinfo {volume} {88}},\ \bibinfo {pages} {043821} (\bibinfo {year} {2013})}\BibitemShut {NoStop}%
\bibitem [{\citenamefont {Sperling}\ \emph {et~al.}(2015)\citenamefont {Sperling}, \citenamefont {Bohmann}, \citenamefont {Vogel}, \citenamefont {Harder}, \citenamefont {Brecht}, \citenamefont {Ansari},\ and\ \citenamefont {Silberhorn}}]{Sperling:PRL:2015}%
  \BibitemOpen
  \bibfield  {author} {\bibinfo {author} {\bibfnamefont {J.}~\bibnamefont {Sperling}}, \bibinfo {author} {\bibfnamefont {M.}~\bibnamefont {Bohmann}}, \bibinfo {author} {\bibfnamefont {W.}~\bibnamefont {Vogel}}, \bibinfo {author} {\bibfnamefont {G.}~\bibnamefont {Harder}}, \bibinfo {author} {\bibfnamefont {B.}~\bibnamefont {Brecht}}, \bibinfo {author} {\bibfnamefont {V.}~\bibnamefont {Ansari}},\ and\ \bibinfo {author} {\bibfnamefont {C.}~\bibnamefont {Silberhorn}},\ }\bibfield  {title} {\bibinfo {title} {Uncovering quantum correlations with time-multiplexed click detection},\ }\href {https://doi.org/10.1103/PhysRevLett.115.023601} {\bibfield  {journal} {\bibinfo  {journal} {Phys. Rev. Lett.}\ }\textbf {\bibinfo {volume} {115}},\ \bibinfo {pages} {023601} (\bibinfo {year} {2015})}\BibitemShut {NoStop}%
\bibitem [{\citenamefont {Serafini}\ \emph {et~al.}(2005)\citenamefont {Serafini}, \citenamefont {Paris}, \citenamefont {Illuminati},\ and\ \citenamefont {Siena}}]{Serafini_2005}%
  \BibitemOpen
  \bibfield  {author} {\bibinfo {author} {\bibfnamefont {A.}~\bibnamefont {Serafini}}, \bibinfo {author} {\bibfnamefont {M.~G.~A.}\ \bibnamefont {Paris}}, \bibinfo {author} {\bibfnamefont {F.}~\bibnamefont {Illuminati}},\ and\ \bibinfo {author} {\bibfnamefont {S.~D.}\ \bibnamefont {Siena}},\ }\bibfield  {title} {\bibinfo {title} {Quantifying decoherence in continuous variable systems},\ }\href {https://doi.org/10.1088/1464-4266/7/4/R01} {\bibfield  {journal} {\bibinfo  {journal} {Journal of Optics B: Quantum and Semiclassical Optics}\ }\textbf {\bibinfo {volume} {7}},\ \bibinfo {pages} {R19} (\bibinfo {year} {2005})}\BibitemShut {NoStop}%
\bibitem [{\citenamefont {Gerry}\ and\ \citenamefont {Grobe}(1995)}]{Gerry:PRA:1995}%
  \BibitemOpen
  \bibfield  {author} {\bibinfo {author} {\bibfnamefont {C.~C.}\ \bibnamefont {Gerry}}\ and\ \bibinfo {author} {\bibfnamefont {R.}~\bibnamefont {Grobe}},\ }\bibfield  {title} {\bibinfo {title} {Nonclassical properties of correlated two-mode schrödinger cat states},\ }\href {https://doi.org/10.1103/PhysRevA.51.1698} {\bibfield  {journal} {\bibinfo  {journal} {Phys. Rev. A}\ }\textbf {\bibinfo {volume} {51}},\ \bibinfo {pages} {1698} (\bibinfo {year} {1995})}\BibitemShut {NoStop}%
\bibitem [{\citenamefont {Hackerv}\ \emph {et~al.}(2019)\citenamefont {Hackerv}, \citenamefont {Welte}, \citenamefont {Daiss}, \citenamefont {Shaukat}, \citenamefont {Ritter}, \citenamefont {Li},\ and\ \citenamefont {Rempe}}]{Hackerv:NP:2019}%
  \BibitemOpen
  \bibfield  {author} {\bibinfo {author} {\bibfnamefont {B.}~\bibnamefont {Hackerv}}, \bibinfo {author} {\bibfnamefont {S.}~\bibnamefont {Welte}}, \bibinfo {author} {\bibfnamefont {S.}~\bibnamefont {Daiss}}, \bibinfo {author} {\bibfnamefont {A.}~\bibnamefont {Shaukat}}, \bibinfo {author} {\bibfnamefont {S.}~\bibnamefont {Ritter}}, \bibinfo {author} {\bibfnamefont {L.}~\bibnamefont {Li}},\ and\ \bibinfo {author} {\bibfnamefont {G.}~\bibnamefont {Rempe}},\ }\bibfield  {title} {\bibinfo {title} {Deterministic creation of entangled atom–light schrödinger-cat states},\ }\href {https://doi.org/10.1038/s41566-018-0339-5} {\bibfield  {journal} {\bibinfo  {journal} {Nature Photon}\ }\textbf {\bibinfo {volume} {13}},\ \bibinfo {pages} {110–115} (\bibinfo {year} {2019})}\BibitemShut {NoStop}%
\bibitem [{\citenamefont {Benenti}\ \emph {et~al.}(2007)\citenamefont {Benenti}, \citenamefont {Casati},\ and\ \citenamefont {Strini}}]{benenti2007tools}%
  \BibitemOpen
  \bibfield  {author} {\bibinfo {author} {\bibfnamefont {G.}~\bibnamefont {Benenti}}, \bibinfo {author} {\bibfnamefont {G.}~\bibnamefont {Casati}},\ and\ \bibinfo {author} {\bibfnamefont {G.}~\bibnamefont {Strini}},\ }\href {https://doi.org/10.1142/10909} {\emph {\bibinfo {title} {Principles of Quantum Computation and Information, Volume II: Basic Tools and Special Topics}}}\ (\bibinfo  {publisher} {World Scientific},\ \bibinfo {year} {2007})\BibitemShut {NoStop}%
\bibitem [{\citenamefont {Weimann}\ \emph {et~al.}(2016)\citenamefont {Weimann}, \citenamefont {Perez-Leija}, \citenamefont {Lebugle}, \citenamefont {Keil}, \citenamefont {Tichy}, \citenamefont {Graefe}, \citenamefont {Heilmann}, \citenamefont {Nolte}, \citenamefont {Moya-Cessa}, \citenamefont {Weihs}, \citenamefont {Christodoulides},\ and\ \citenamefont {Szameit}}]{Weimann2016}%
  \BibitemOpen
  \bibfield  {author} {\bibinfo {author} {\bibfnamefont {S.}~\bibnamefont {Weimann}}, \bibinfo {author} {\bibfnamefont {A.}~\bibnamefont {Perez-Leija}}, \bibinfo {author} {\bibfnamefont {M.}~\bibnamefont {Lebugle}}, \bibinfo {author} {\bibfnamefont {R.}~\bibnamefont {Keil}}, \bibinfo {author} {\bibfnamefont {M.}~\bibnamefont {Tichy}}, \bibinfo {author} {\bibfnamefont {M.}~\bibnamefont {Graefe}}, \bibinfo {author} {\bibfnamefont {R.}~\bibnamefont {Heilmann}}, \bibinfo {author} {\bibfnamefont {S.}~\bibnamefont {Nolte}}, \bibinfo {author} {\bibfnamefont {H.}~\bibnamefont {Moya-Cessa}}, \bibinfo {author} {\bibfnamefont {G.}~\bibnamefont {Weihs}}, \bibinfo {author} {\bibfnamefont {D.~N.}\ \bibnamefont {Christodoulides}},\ and\ \bibinfo {author} {\bibfnamefont {A.}~\bibnamefont {Szameit}},\ }\bibfield  {title} {\bibinfo {title} {Implementation of quantum and classical discrete fractional fourier transforms},\ }\href {https://doi.org/10.1038/ncomms11027} {\bibfield  {journal} {\bibinfo  {journal} {Nat. Commun.}\
  }\textbf {\bibinfo {volume} {7}},\ \bibinfo {pages} {11027} (\bibinfo {year} {2016})}\BibitemShut {NoStop}%
\bibitem [{\citenamefont {Selim}\ \emph {et~al.}(2025)\citenamefont {Selim}, \citenamefont {Ehrhardt}, \citenamefont {Ding}, \citenamefont {Dinani}, \citenamefont {Zhong}, \citenamefont {Perez‐Leija}, \citenamefont {Şahin K.~Özdemir}, \citenamefont {Heinrich}, \citenamefont {Szameit}, \citenamefont {Christodoulides},\ and\ \citenamefont {Khajavikhan}}]{Selim:Science:2025}%
  \BibitemOpen
  \bibfield  {author} {\bibinfo {author} {\bibfnamefont {M.~A.}\ \bibnamefont {Selim}}, \bibinfo {author} {\bibfnamefont {M.}~\bibnamefont {Ehrhardt}}, \bibinfo {author} {\bibfnamefont {Y.}~\bibnamefont {Ding}}, \bibinfo {author} {\bibfnamefont {H.~M.}\ \bibnamefont {Dinani}}, \bibinfo {author} {\bibfnamefont {Q.}~\bibnamefont {Zhong}}, \bibinfo {author} {\bibfnamefont {A.}~\bibnamefont {Perez‐Leija}}, \bibinfo {author} {\bibnamefont {Şahin K.~Özdemir}}, \bibinfo {author} {\bibfnamefont {M.}~\bibnamefont {Heinrich}}, \bibinfo {author} {\bibfnamefont {A.}~\bibnamefont {Szameit}}, \bibinfo {author} {\bibfnamefont {D.~N.}\ \bibnamefont {Christodoulides}},\ and\ \bibinfo {author} {\bibfnamefont {M.}~\bibnamefont {Khajavikhan}},\ }\bibfield  {title} {\bibinfo {title} {Selective filtering of photonic quantum entanglement via anti-parity-time symmetry},\ }\href {https://doi.org/10.1126/science.adu3777} {\bibfield  {journal} {\bibinfo  {journal} {Science}\ }\textbf {\bibinfo {volume} {387}},\ \bibinfo {pages}
  {1424} (\bibinfo {year} {2025})}\BibitemShut {NoStop}%
\bibitem [{\citenamefont {Politi}\ \emph {et~al.}(2008)\citenamefont {Politi}, \citenamefont {Cryan}, \citenamefont {Rarity}, \citenamefont {Yu},\ and\ \citenamefont {O'Brien}}]{Politi:Science}%
  \BibitemOpen
  \bibfield  {author} {\bibinfo {author} {\bibfnamefont {A.}~\bibnamefont {Politi}}, \bibinfo {author} {\bibfnamefont {M.~J.}\ \bibnamefont {Cryan}}, \bibinfo {author} {\bibfnamefont {J.~G.}\ \bibnamefont {Rarity}}, \bibinfo {author} {\bibfnamefont {S.}~\bibnamefont {Yu}},\ and\ \bibinfo {author} {\bibfnamefont {J.~L.}\ \bibnamefont {O'Brien}},\ }\bibfield  {title} {\bibinfo {title} {Silica-on-silicon waveguide quantum circuits},\ }\href {https://doi.org/10.1126/science.1155441} {\bibfield  {journal} {\bibinfo  {journal} {Science}\ }\textbf {\bibinfo {volume} {320}},\ \bibinfo {pages} {646} (\bibinfo {year} {2008})},\ \Eprint {https://arxiv.org/abs/https://www.science.org/doi/pdf/10.1126/science.1155441} {https://www.science.org/doi/pdf/10.1126/science.1155441} \BibitemShut {NoStop}%
\bibitem [{\citenamefont {Provazník}\ \emph {et~al.}(2020)\citenamefont {Provazník}, \citenamefont {Lachman}, \citenamefont {Filip},\ and\ \citenamefont {Marek}}]{Provazník:OE:2013}%
  \BibitemOpen
  \bibfield  {author} {\bibinfo {author} {\bibfnamefont {J.}~\bibnamefont {Provazník}}, \bibinfo {author} {\bibfnamefont {L.}~\bibnamefont {Lachman}}, \bibinfo {author} {\bibfnamefont {R.}~\bibnamefont {Filip}},\ and\ \bibinfo {author} {\bibfnamefont {P.}~\bibnamefont {Marek}},\ }\bibfield  {title} {\bibinfo {title} {Benchmarking photon number resolving detectors},\ }\href {https://doi.org/10.1364/OE.389619} {\bibfield  {journal} {\bibinfo  {journal} {Opt. Express}\ }\textbf {\bibinfo {volume} {28}},\ \bibinfo {pages} {14839} (\bibinfo {year} {2020})}\BibitemShut {NoStop}%
\bibitem [{\citenamefont {Anaya-Contreras}\ \emph {et~al.}(2021)\citenamefont {Anaya-Contreras}, \citenamefont {Zúñiga-Segundo}, \citenamefont {Perez-Leija}, \citenamefont {de~J~Leon-Montiel},\ and\ \citenamefont {Moya-Cessa}}]{Anaya-Contreras_2021}%
  \BibitemOpen
  \bibfield  {author} {\bibinfo {author} {\bibfnamefont {J.~A.}\ \bibnamefont {Anaya-Contreras}}, \bibinfo {author} {\bibfnamefont {A.}~\bibnamefont {Zúñiga-Segundo}}, \bibinfo {author} {\bibfnamefont {A.}~\bibnamefont {Perez-Leija}}, \bibinfo {author} {\bibfnamefont {R.}~\bibnamefont {de~J~Leon-Montiel}},\ and\ \bibinfo {author} {\bibfnamefont {H.~M.}\ \bibnamefont {Moya-Cessa}},\ }\bibfield  {title} {\bibinfo {title} {Multiphoton processes via conditional measurements in the two-field interaction},\ }\href {https://doi.org/10.1088/2040-8986/ac182f} {\bibfield  {journal} {\bibinfo  {journal} {J. Opt.}\ }\textbf {\bibinfo {volume} {23}},\ \bibinfo {pages} {095201} (\bibinfo {year} {2021})}\BibitemShut {NoStop}%
\bibitem [{\citenamefont {Datta}\ \emph {et~al.}(2024)\citenamefont {Datta}, \citenamefont {Perez-Leija},\ and\ \citenamefont {Busch}}]{Datta:24}%
  \BibitemOpen
  \bibfield  {author} {\bibinfo {author} {\bibfnamefont {A.~M.}\ \bibnamefont {Datta}}, \bibinfo {author} {\bibfnamefont {A.}~\bibnamefont {Perez-Leija}},\ and\ \bibinfo {author} {\bibfnamefont {K.}~\bibnamefont {Busch}},\ }\bibfield  {title} {\bibinfo {title} {Tailoring the nonclassicality of light states via mode detuning in waveguide beam splitters},\ }\href {https://doi.org/10.1364/JOSAB.521859} {\bibfield  {journal} {\bibinfo  {journal} {J. Opt. Soc. Am. B}\ }\textbf {\bibinfo {volume} {41}},\ \bibinfo {pages} {1557} (\bibinfo {year} {2024})}\BibitemShut {NoStop}%
\bibitem [{\citenamefont {Wang}\ \emph {et~al.}(2020)\citenamefont {Wang}, \citenamefont {Sciarrino}, \citenamefont {Laing},\ and\ \citenamefont {Thompson}}]{Wang:NP:2020}%
  \BibitemOpen
  \bibfield  {author} {\bibinfo {author} {\bibfnamefont {J.}~\bibnamefont {Wang}}, \bibinfo {author} {\bibfnamefont {F.}~\bibnamefont {Sciarrino}}, \bibinfo {author} {\bibfnamefont {A.}~\bibnamefont {Laing}},\ and\ \bibinfo {author} {\bibfnamefont {M.~G.}\ \bibnamefont {Thompson}},\ }\bibfield  {title} {\bibinfo {title} {Integrated photonic quantum technologies},\ }\href {https://doi.org/10.1038/s41566-019-0532-1} {\bibfield  {journal} {\bibinfo  {journal} {Nat. Photonics}\ }\textbf {\bibinfo {volume} {14}},\ \bibinfo {pages} {273–284} (\bibinfo {year} {2020})}\BibitemShut {NoStop}%
\bibitem [{\citenamefont {Labont\'e}\ \emph {et~al.}(2024)\citenamefont {Labont\'e}, \citenamefont {Alibart}, \citenamefont {D'Auria}, \citenamefont {Doutre}, \citenamefont {Etesse}, \citenamefont {Sauder}, \citenamefont {Martin}, \citenamefont {Picholle},\ and\ \citenamefont {Tanzilli}}]{Laurent:PRXQ:2024}%
  \BibitemOpen
  \bibfield  {author} {\bibinfo {author} {\bibfnamefont {L.}~\bibnamefont {Labont\'e}}, \bibinfo {author} {\bibfnamefont {O.}~\bibnamefont {Alibart}}, \bibinfo {author} {\bibfnamefont {V.}~\bibnamefont {D'Auria}}, \bibinfo {author} {\bibfnamefont {F.}~\bibnamefont {Doutre}}, \bibinfo {author} {\bibfnamefont {J.}~\bibnamefont {Etesse}}, \bibinfo {author} {\bibfnamefont {G.}~\bibnamefont {Sauder}}, \bibinfo {author} {\bibfnamefont {A.}~\bibnamefont {Martin}}, \bibinfo {author} {\bibfnamefont {E.}~\bibnamefont {Picholle}},\ and\ \bibinfo {author} {\bibfnamefont {S.}~\bibnamefont {Tanzilli}},\ }\bibfield  {title} {\bibinfo {title} {Integrated photonics for quantum communications and metrology},\ }\href {https://doi.org/10.1103/PRXQuantum.5.010101} {\bibfield  {journal} {\bibinfo  {journal} {PRX Quantum}\ }\textbf {\bibinfo {volume} {5}},\ \bibinfo {pages} {010101} (\bibinfo {year} {2024})}\BibitemShut {NoStop}%
\bibitem [{\citenamefont {Wang}\ \emph {et~al.}(2025)\citenamefont {Wang}, \citenamefont {Chai}, \citenamefont {Cao}, \citenamefont {Chen}, \citenamefont {Liang},\ and\ \citenamefont {Peng}}]{Wang2025}%
  \BibitemOpen
  \bibfield  {author} {\bibinfo {author} {\bibfnamefont {L.}~\bibnamefont {Wang}}, \bibinfo {author} {\bibfnamefont {G.}~\bibnamefont {Chai}}, \bibinfo {author} {\bibfnamefont {Z.}~\bibnamefont {Cao}}, \bibinfo {author} {\bibfnamefont {X.}~\bibnamefont {Chen}}, \bibinfo {author} {\bibfnamefont {K.}~\bibnamefont {Liang}},\ and\ \bibinfo {author} {\bibfnamefont {J.}~\bibnamefont {Peng}},\ }\bibfield  {title} {\bibinfo {title} {Quantum purification for coherent states and its application},\ }\href {https://doi.org/10.1007/s11433-024-2533-4} {\bibfield  {journal} {\bibinfo  {journal} {Sci. China Phys. Mech. Astron.}\ }\textbf {\bibinfo {volume} {68}},\ \bibinfo {pages} {220313} (\bibinfo {year} {2025})}\BibitemShut {NoStop}%
\bibitem [{\citenamefont {Zhao}\ \emph {et~al.}(2022)\citenamefont {Zhao}, \citenamefont {Li}, \citenamefont {Xu}, \citenamefont {Huang}, \citenamefont {Wang},\ and\ \citenamefont {Zeng}}]{Zhao:22}%
  \BibitemOpen
  \bibfield  {author} {\bibinfo {author} {\bibfnamefont {H.}~\bibnamefont {Zhao}}, \bibinfo {author} {\bibfnamefont {H.}~\bibnamefont {Li}}, \bibinfo {author} {\bibfnamefont {Y.}~\bibnamefont {Xu}}, \bibinfo {author} {\bibfnamefont {P.}~\bibnamefont {Huang}}, \bibinfo {author} {\bibfnamefont {T.}~\bibnamefont {Wang}},\ and\ \bibinfo {author} {\bibfnamefont {G.}~\bibnamefont {Zeng}},\ }\bibfield  {title} {\bibinfo {title} {Simple continuous-variable quantum key distribution scheme using a sagnac-based gaussian modulator},\ }\href {https://doi.org/10.1364/OL.458443} {\bibfield  {journal} {\bibinfo  {journal} {Opt. Lett.}\ }\textbf {\bibinfo {volume} {47}},\ \bibinfo {pages} {2939} (\bibinfo {year} {2022})}\BibitemShut {NoStop}%
\bibitem [{\citenamefont {Spring}\ \emph {et~al.}(2013)\citenamefont {Spring}, \citenamefont {Metcalf}, \citenamefont {Humphreys}, \citenamefont {Kolthammer}, \citenamefont {Jin}, \citenamefont {Barbieri}, \citenamefont {Datta}, \citenamefont {Thomas-Peter}, \citenamefont {Langford}, \citenamefont {Kundys}, \citenamefont {Gates}, \citenamefont {Smith}, \citenamefont {Smith},\ and\ \citenamefont {Walmsley}}]{Spring:Science:2013}%
  \BibitemOpen
  \bibfield  {author} {\bibinfo {author} {\bibfnamefont {J.~B.}\ \bibnamefont {Spring}}, \bibinfo {author} {\bibfnamefont {B.~J.}\ \bibnamefont {Metcalf}}, \bibinfo {author} {\bibfnamefont {P.~C.}\ \bibnamefont {Humphreys}}, \bibinfo {author} {\bibfnamefont {W.~S.}\ \bibnamefont {Kolthammer}}, \bibinfo {author} {\bibfnamefont {X.-M.}\ \bibnamefont {Jin}}, \bibinfo {author} {\bibfnamefont {M.}~\bibnamefont {Barbieri}}, \bibinfo {author} {\bibfnamefont {A.}~\bibnamefont {Datta}}, \bibinfo {author} {\bibfnamefont {N.}~\bibnamefont {Thomas-Peter}}, \bibinfo {author} {\bibfnamefont {N.~K.}\ \bibnamefont {Langford}}, \bibinfo {author} {\bibfnamefont {D.}~\bibnamefont {Kundys}}, \bibinfo {author} {\bibfnamefont {J.~C.}\ \bibnamefont {Gates}}, \bibinfo {author} {\bibfnamefont {B.~J.}\ \bibnamefont {Smith}}, \bibinfo {author} {\bibfnamefont {P.~G.~R.}\ \bibnamefont {Smith}},\ and\ \bibinfo {author} {\bibfnamefont {I.~A.}\ \bibnamefont {Walmsley}},\ }\bibfield  {title} {\bibinfo {title} {Boson sampling on a photonic
  chip},\ }\href {https://doi.org/10.1126/science.1231692} {\bibfield  {journal} {\bibinfo  {journal} {Science}\ }\textbf {\bibinfo {volume} {339}},\ \bibinfo {pages} {798} (\bibinfo {year} {2013})}\BibitemShut {NoStop}%
\end{thebibliography}%

\end{document}